\documentclass[12pt, preprint]{aastex}

\newcommand\arcpt{${{\lower3pt\hbox{$^{\prime\prime}$}}\atop{\raise4pt\hbox{.}}}$}

\slugcomment{to appear in the {\it Astronomical Journal}}

\shorttitle{New High Proper Motion Stars}
\shortauthors{Subasavage et al.}

\begin{document}

\title{The Solar Neighborhood XV: \\ Discovery of New High Proper
Motion Stars \\ with $\mu$ $\ge$ 0$\farcs$4 yr$^{-1}$ between
Declinations $-$47$^\circ$ and 00$^\circ$ }

\author{John P. Subasavage, Todd J. Henry}

\affil{Georgia State University, Atlanta, GA 30302-4106}

\author{Nigel C. Hambly}

\affil{Institute for Astronomy, University of Edinburgh \\ Royal
Observatory, Blackford Hill, Edinburgh, EH9~3HJ, Scotland, UK}

\author{Misty A. Brown, Wei-Chun Jao, and Charlie T. Finch}

\affil{Georgia State University, Atlanta, GA 30302-4106}

\email{subasavage@chara.gsu.edu}

%%%%%%%%%%%%%%%%%%%%%%%%%%%%%%%%%%%%%%%%%%%%%%%%%%%%%%%%%%%%%%%%%%%%%%%%%%%%%%
% {Abstract}
%%%%%%%%%%%%%%%%%%%%%%%%%%%%%%%%%%%%%%%%%%%%%%%%%%%%%%%%%%%%%%%%%%%%%%%%%%%%%%

\begin{abstract}

We report the discovery of 152 new high proper motion systems ($\mu$
$\ge$ 0$\farcs$4 yr$^{-1}$) in the southern sky ($\delta$ $=$
$-$47$^\circ$ to 00$^\circ$) brighter than UKST plate $R_{59F}$ $=$
16.5 via our SuperCOSMOS-RECONS (SCR) search.  This paper complements
Paper XII in The Solar Neighborhood series, which covered the region
from $\delta$ $=$ $-$90$^\circ$ to $-$47$^\circ$ and discussed all 147
new systems from the southernmost phase of the search.  Among the
total of 299 systems from both papers, there are 148 (71 in Paper XII,
77 in this paper) new systems moving faster than 0$\farcs$5 yr$^{-1}$
that are additions to the classic ``LHS'' (Luyten Half Second) sample.
These constitute an 8\% increase in the sample of all stellar systems
with $\mu$ $\ge$ 0$\farcs$5 yr$^{-1}$ in the southern sky.

As in Paper XII, distance estimates are provided for the systems
reported here based upon a combination of photographic plate
magnitudes and 2MASS photometry, assuming all stars are on the main
sequence.  Two SCR systems from the portion of the sky included in
this paper are anticipated to be within 10 pc, and an additional 23
are within 25 pc.  In total, the results presented in Paper XII and
here for this SCR sweep of the entire southern sky include five new
systems within 10 pc and 38 more between 10 and 25 pc.  The largest
number of nearby systems have been found in the slowest proper motion
bin, 0$\farcs$6 yr$^{-1}$ $>$ $\mu$ $\ge$ 0$\farcs$4 yr$^{-1}$, indicating that
there may be a large population of low proper motion systems very near
the Sun.

\end{abstract}


%%%%%%%%%%%%%%%%%%%%%%%%%%%%%%%%%%%%%%%%%%%%%%%%%%%%%%%%%%%%%%%%%%%%%%%%%%%%%%
\section{Introduction}
%%%%%%%%%%%%%%%%%%%%%%%%%%%%%%%%%%%%%%%%%%%%%%%%%%%%%%%%%%%%%%%%%%%%%%%%%%%%%%

In this paper we continue our reconnaissance of the solar neighborhood
by reporting results of the second phase of our SuperCOSMOS-RECONS
(SCR) search for new high proper motion (HPM) objects in the southern
sky.  In Paper XII of The Solar Neighborhood series
\citep{2005AJ....129..413S}, we presented results for the portion of
the sky between declinations $-$90$^\circ$ and $-$47$^\circ$.  In this
paper we cover the region from declinations $-$47$^\circ$ to
00$^\circ$.

Recently, the northern sky has been searched systematically for HPM
objects by L{\' e}pine et al. \citeyearpar{2002AJ....124.1190L,
2003AJ....126..921L} utilizing SUPERBLINK, which has been particularly
successful at filling in the distribution of HPM objects along the
Galactic plane.  Their latest compendium \citep{2005AJ....129.1483L}
includes 61977 HPM objects with $\mu$ $\ge$ 0$\farcs$15 yr$^{-1}$.
The total number of objects is roughly double what was found during
the pioneering days of \citet{1971lpms.book.....G,1978LowOB...8...89G}
and \citet{1979lccs.book.....L,1995yCat.1098....0L}.  Historically,
however, the southern sky has been investigated with limited depth for
HPM objects, which are notoriously good nearby star candidates.

Since the Giclas and Luyten efforts, significant numbers of HPM stars
have been found, primarily because of the advent of large databases
holding data from digitized photographic plates.  Some of the new
discoveries are remarkably nearby, including SO 0253+1652
\citep{2003ApJ...589L..51T}, DENIS 1048-3956
\citep{2001A&A...366L..13D}, and SCR 1845-6357
\citep{2004AJ....128..437H}, which have distances estimated to be 3.7,
4.5, and 4.6 pc, respectively \citep{2004AJ....128.2460H}.  Because of
their proximity, these three red dwarfs are high priority targets in
our trigonometric parallax program, CTIOPI (Cerro Tololo
Inter-American Observatory Parallax Investigation) being carried out
in Chile \citep{2005AJ....129.1954J, 2005ASPC...00..000H}.  As is the
case for all of the recent HPM surveys, the new SCR discoveries are
primarily red dwarfs that are underrepresented in current compendia of
solar neighborhood membership lists because of their intrinsic
faintness \citep{1997AJ....114..388H}.  Additional smatterings of
white dwarfs and subdwarfs are also found in the SCR and other HPM
searches.

The classic work of Giclas and Luyten has been complemented by the
recent HPM surveys summarized in Table~\ref{pm-surveys}, each of which
has revealed important new HPM objects (the machine selected catalog
of 11289 objects generated by \citet{2004A&A...421..763P} remains
difficult to assess because our initial checks indicate that many of
the objects are previously known and several are not real HPM
sources).  We compare the number of new discoveries from each survey
to the Luyten Half Second Catalogue \citep[][hereafter
LHS]{1979lccs.book.....L}, which included 3602 objects with proper
motions in excess of 0$\farcs$5 yr$^{-1}$, and still accounts for 87\%
of all known such objects.  This reveals the enormous impact of the
work by Giclas and Luyten, who carried out their surveys in times
before massive computer searches of digitized photographic plates were
possible.

%%%%%%%%%%%%%%%%%%%%%%%%%%%%%%%%%%%%%%%%%%%%%%%%%%%%%%%%%%%%%%%%%%%%%%%%%%%%%%
\section {Search Methodology}
%%%%%%%%%%%%%%%%%%%%%%%%%%%%%%%%%%%%%%%%%%%%%%%%%%%%%%%%%%%%%%%%%%%%%%%%%%%%%%

To reveal new HPM objects, we mine the SuperCOSMOS database developed
and maintained at the Royal Observatory in Edinburgh, Scotland.
Papers VIII \citep{2004AJ....128..437H}, X
\citep{2004AJ....128.2460H}, and XII \citep{2005AJ....129..413S} in
this series include previous discoveries from our SCR survey.  The
search techniques utilized here are identical to those in
\citet{2004AJ....128..437H}, where a full discussion can be found.
Briefly, we utilize the SuperCOSMOS Sky Survey to reveal previously
unknown HPM objects using a combination of astrometric and photometric
information from the four photographic plates available ($B_J$,
$ESO-$$R$, $R_{59F}$, and $I_{IVN}$) in the southern sky.

In Paper XII, 1424 candidate objects were found having
10$\farcs$0 yr$^{-1}$ $>$ $\mu$ $\ge$ 0$\farcs$4 yr$^{-1}$ and brighter than
$R_{59F}$ $=$ 16.5.  In this portion of the search, we have found an
additional 3879 candidates meeting the same criteria.  The combined
coverage of the two portions of the survey includes 46\% of the entire
sky, and 92\% of the southern sky, where we have searched from the
south celestial pole at $\delta$ $=$ $-$90$^\circ$ to a northern
cutoff at precisely $\delta$ $=$ 00$^\circ$ (even though the plates
typically extend to $\delta$ $=$ $+$3.0$^\circ$).
Figure~\ref{pltfields} is a map of the sky coverage, including the 894
plate fields in the southern sky.  Seventy-one fields have not been
searched because of crowding near the Galactic plane or Magellanic
Clouds, or a limited spread in epochs for available plates.

The vetting of candidates, which includes checks of proper motions,
magnitudes, colors, and image ellipticities, as well as inspection by
eye, was described in detail in Paper XII.  For each candidate that
appeared to be a real object, coordinates were carefully cross-checked
with the NLTT catalog, the SIMBAD database, and recent HPM
publications (see Table~\ref{pm-surveys}) to see if it was a known
object.  If the coordinates agreed to within a few arcminutes and the
magnitudes and proper motions were consistent, the detection was
considered previously known.  In a few cases, the coordinates and
proper motions agreed well, but the magnitudes did not.  These objects
were revealed to be new wide common proper motion companions to
previously known proper motion stars --- four new companions are
reported here (see $\S$ 5.5).  Overall, the hit rate for new HPM
objects decreases with increasing proper motion because reliable
source association between different epochs is more difficult for
fast-moving sources.  For objects with 10$\farcs$0 yr$^{-1}$ $>$ $\mu$
$\ge$ 1$\farcs$0 yr$^{-1}$, only 10\% turn out to be real, whereas
87\% of objects detected with 1$\farcs$0 yr$^{-1}$ $>$ $\mu$ $\ge$
0$\farcs$4 yr$^{-1}$ are real.  These fractions include both new and
known objects.

The final count of real, distinct, new systems with 10$\farcs$0
yr$^{-1}$ $>$ $\mu$ $\ge$ 0$\farcs$4 yr$^{-1}$ and brighter than
$R_{59F}$ $=$ 16.5 found between $\delta$ $=$ $-$47$^\circ$ and
00$^\circ$ is 152.  Finder charts are given at the end of this paper
in Figure~\ref{finders} for all of the new systems reported here, as
well as for the four new wide companions.  We continue using our
naming convention, ``SCR'' for objects discovered during the survey.

%%%%%%%%%%%%%%%%%%%%%%%%%%%%%%%%%%%%%%%%%%%%%%%%%%%%%%%%%%%%%%%%%%%%%%%%%%%%%%
\section {Comparison to Previous Proper Motion Surveys}
%%%%%%%%%%%%%%%%%%%%%%%%%%%%%%%%%%%%%%%%%%%%%%%%%%%%%%%%%%%%%%%%%%%%%%%%%%%%%%

A primary goal of the SCR effort is to further complete the LHS
Catalogue for stars with $\mu$ $\ge$ 0$\farcs$5 yr$^{-1}$.  Our
extension of the cutoff to $\mu$ $\ge$ 0$\farcs$4 yr$^{-1}$ in this
survey is to ensure that no known LHS stars were missed due to proper
motion measurement errors for objects very near the 0$\farcs$5
yr$^{-1}$ limit.  Of the 299 new SCR systems found to date, 148 have
$\mu$ $\ge$ 0$\farcs$5 yr$^{-1}$.  Figure~\ref{surveys} provides a map
of new systems with $\mu$ $\ge$ 0$\farcs$5 yr$^{-1}$, including the
148 total SCR systems (solid triangles representing 150 individual
objects) that are additions to the LHS sample.

\citet{2002AJ....124.1190L,2003AJ....126..921L,2005AJ....129.1483L}
have completed work on the entire northern sky using SUPERBLINK and
are so far the most productive survey for revealing new objects with
$\mu$ $\ge$ 0$\farcs$5 yr$^{-1}$.  Although we avoid the Galactic plane and
Magellanic Clouds, the SCR survey has the most uniform sky coverage
for southern hemisphere searches, and is consequently the most
productive survey in the south.  There are undoubtedly objects
remaining to be discovered along the crowded Galactic plane in the
south --- the gaps in Figure~\ref{surveys} typically match the regions
not searched (white spaces in Figure~\ref{pltfields}).  The same is
true of the LMC ($\alpha$ $=$ 05:30H, $\delta$ $=$ $-$68$^\circ$) and
SMC ($\alpha$ $=$ 01:00H, $\delta$ $=$ $-$73$^\circ$) regions.

There are 1462 LHS stars in the LHS Catalogue brighter than a
photographic R magnitude (R$_{pg}$) of 16.5 in the southern sky.  Of
the 1152 known LHS stars in the southern sky with 10.0 $<$ R$_{pg}$
$<$ 16.5, we have recovered 1032 (90\%).  We recover only 234 of 310
(75\%) of LHS stars brighter than R $=$ 10.0 because the search is
somewhat less sensitive to bright objects that are saturated in the
photographic emulsions.  This recovery rate is somewhat less
successful for stars moving faster than 1$\farcs$0 yr$^{-1}$ (199 of
251, 79\%) than for stars with $\mu =$ 0$\farcs$5--1$\farcs$0
yr$^{-1}$ (1067 of 1211, 88\%).

%%%%%%%%%%%%%%%%%%%%%%%%%%%%%%%%%%%%%%%%%%%%%%%%%%%%%%%%%%%%%%%%%%%%%%%%%%%%%%
\section {Data}
%%%%%%%%%%%%%%%%%%%%%%%%%%%%%%%%%%%%%%%%%%%%%%%%%%%%%%%%%%%%%%%%%%%%%%%%%%%%%%

As in Paper XII, coordinates, proper motions, and plate magnitudes
have been extracted from SuperCOSMOS for the new HPM systems.  These
data are listed in Table~\ref{scr-tbl} for objects in the new portion
of sky searched here.  Coordinates are for epoch and equinox J2000.
Errors in the coordinates are typically $\pm$ 0$\farcs$3, and errors
in the proper motions are given.  Errors in position angle are usually
$\pm$ 0.1$^\circ$.  Photometric magnitudes are given for three sets of
plates --- $B_J$, $R_{59F}$, and $I_{IVN}$.  Magnitude errors are
$\sim$0.3 mag or better for $m$ $>$ 15 and actually get larger at
brighter magnitudes due to systematic errors
\citep{2001MNRAS.326.1295H}.  A few plate magnitude values are missing
because of blending problems that preclude accurate magnitude
determinations.

Infrared photometry has been used to extend the color baseline, which
allows more accurate photometric distance estimates for red dwarfs and
permits a reliable separation of the white and red dwarfs.  The
infrared $JHK_s$ photometry has been extracted from 2MASS via Aladin.
Each SCR object has been identified by eye to ensure that no extracted
magnitudes are in error.  In nearly every case, the errors are smaller
than 0.03 mag.  Exceptions include objects with $J$ $>$ 15, $H$ $>$
14.5, and $K_s$ $>$ 14, where the errors are 0.05 mag or greater.  In
one case, SCR 1246-1236, the error is null for $K_s$, and the value is
therefore unreliable.

%%%%%%%%%%%%%%%%%%%%%%%%%%%%%%%%%%%%%%%%%%%%%%%%%%%%%%%%%%%%%%%%%%%%%%%%%%%%%%
\section {Analysis}
%%%%%%%%%%%%%%%%%%%%%%%%%%%%%%%%%%%%%%%%%%%%%%%%%%%%%%%%%%%%%%%%%%%%%%%%%%%%%%
%%%%%%%%%%%%%%%%%%%%%%%%%%%%%%%%%%%%%%%%%%%%%%%%%%%%%%%%%%%%%%%%%%%%%%%%%%%%%%
\subsection {Color-Magnitude Diagram}
%%%%%%%%%%%%%%%%%%%%%%%%%%%%%%%%%%%%%%%%%%%%%%%%%%%%%%%%%%%%%%%%%%%%%%%%%%%%%%

Illustrated in Figure~\ref{colmag} is a color-apparent magnitude
diagram which compares new SCR objects (split into large solid points
representing objects with $\mu$ $\ge$ 0$\farcs$5 yr$^{-1}$ and small
solid points representing objects with 0$\farcs$5 yr$^{-1}$ $>$ $\mu$
$\ge$ 0$\farcs$4 yr$^{-1}$) to previously known stars having $\mu$
$\ge$ 0$\farcs$5 yr$^{-1}$.  As in Paper XII, the SCR discoveries are
generally fainter and redder than the bulk of the known stars (note
that the open/closed symbols have been exchanged relative to Paper XII
for clarity in this figure).  However, the portion of sky searched in
this paper is much larger (2.2 times the area) than that targeted in
Paper XII and overlaps the regions searched by Giclas and Luyten.
This results in two important differences from the similar Figure 4 in
Paper XII.  First, there are far more known objects shown here than in
the deep southern sky.  Second, because Luyten's Bruce Proper Motion
survey for the sky south of $\delta$ $=$ $-$33$^\circ$ had a
relatively bright limit ($m_{pg}\sim$15.5) and lacked red plates,
there were very few known faint stars in the complementary figure of
Paper XII.  Here there are many faint stars from the overlap region
between $\delta$ $=$ $-$33$^\circ$ to 00$^\circ$ where earlier
searches did have faint limits.  Note also that there are 43 white
dwarfs here (39 known and 4 new candidates), whereas there were 16 in
Paper XII, entirely consistent with a uniform white dwarf distribution
on the sky given the ratio of sky areas searched.

We have not found any additional extremely red objects like SCR
1845-6357 [$R_{59F}-J$ $=$ 7.79, $R_{59F}$ $=$ 16.33,
\citet{2004AJ....128..437H}; $V-K_s$ $=$ 8.89, M8.5V, distance
estimate 4.6 pc, \citet{2004AJ....128.2460H}; trigonometric parallax
282 $\pm$ 23 mas, \citet{2005AJ....129..409D}] in this portion of the
sky.  Instead, the remarkable find is SCR 0640-0552 with $R_{59F}-J$
$=$ 1.95, $R_{59F}$ $=$ 8.79, and a distance estimated here to be 8.5
pc, assuming it is single (see $\S$ 5.5).  That such a bright object
remained unidentified until this survey is surprising, indicating yet
again that there may be very nearby, relatively bright stars that have
not yet been found.  As noticed in Paper XII, there is not a
significant drop in the number of objects at our adopted faint limit,
so there is likely to remain a large population of fainter objects yet
to be discovered in the SuperCOSMOS data.  Moreover, the population of
objects with $\mu$ $<$ 0$\farcs$5 yr$^{-1}$ has barely been
investigated, although a portion of the previously known objects in
the NLTT have been examined \citep{2004AJ....128..463R}.

%%%%%%%%%%%%%%%%%%%%%%%%%%%%%%%%%%%%%%%%%%%%%%%%%%%%%%%%%%%%%%%%%%%%%%%%%%%%%%
\subsection {Reduced Proper Motion Diagram}
%%%%%%%%%%%%%%%%%%%%%%%%%%%%%%%%%%%%%%%%%%%%%%%%%%%%%%%%%%%%%%%%%%%%%%%%%%%%%%

Shown in Figure~\ref{redpromo} is the reduced proper motion (RPM)
diagram for objects from the region of sky reported here.  The RPM
diagram is used to separate effectively the white dwarfs from main
sequence stars, as well as to assist in identifying subdwarfs.  The
assumption is, of course, that proper motion is directly related to
distance.  A complementary plot can be found in Figure 5 of Paper XII
for the southernmost portion of the SCR survey.

As in Paper XII it is apparent that most of the new SCR stars are main
sequence red dwarfs, while there is a substantial sample of new
subdwarf candidates --- note the bifurcated population of round points
running from the upper left to lower right in Figure~\ref{redpromo};
the area just above the dotted line maps out the subdwarf region.  The
dotted line represents a somewhat arbitrary boundary between the
subdwarfs and white dwarfs.  There are four clear white dwarf
candidates, two of which are confirmed white dwarfs (solid triangles)
--- for one we have a confirmation spectrum, and the other is a common
proper motion companion to a star of known distance.

%%%%%%%%%%%%%%%%%%%%%%%%%%%%%%%%%%%%%%%%%%%%%%%%%%%%%%%%%%%%%%%%%%%%%%%%%%%%%%
\subsection {Red Dwarfs and Subdwarfs}
%%%%%%%%%%%%%%%%%%%%%%%%%%%%%%%%%%%%%%%%%%%%%%%%%%%%%%%%%%%%%%%%%%%%%%%%%%%%%%

The combination of the $B_J$, $R_{59F}$, and $I_{IVN}$ plate
magnitudes and $JHK_s$ photometry from 2MASS allows us to estimate
distances to main sequence stars, as described in
\citet{2004AJ....128..437H}.  The six magnitudes provide 15
color--M$_K$ combinations, 11 of which can be used to estimate
individual distances ($JHK_s$--only colors are not used because of
limited color discrimination, and $B_J - R_{59F}$ is not sensitive to
absolute magnitude for cooler dwarfs).  The relations assume that the
objects are single, main sequence, dwarfs of types $\sim$K0V to M9V.
The distances have an average error of 26\%, determined by running the
RECONS 10 pc sample of single red dwarfs with known distances through
the suite of photometric distance relations.  For our investigation,
the most interesting stars are those that are potentially nearby,
specifically within the volumes defined by the RECONS 10 pc and the
CNS \citep[Catalog of Nearby Stars,][]{1991adc..rept.....G} and NStars
(Nearby Stars) samples (horizons at 25 pc).

Distance estimates for the 152 new systems in the portion of the sky
searched here are given in final column of Table~\ref{scr-tbl}.  If
only one color was available for a distance estimate, no distance is
reported because the single color falling within the bounds of the
relations is possibly aberrant.  Two of the systems, SCR 0640-0552 and
SCR 0740-4257, are new candidates for the 10 pc sample, and an
additional 23 systems are estimated to be between 10 and 25 pc.

There are 9 SCR stars having distance estimates in excess of 200 pc
(excluding white dwarf candidates), and several more with distance
estimates very close to 200 pc, most of which fall blueward of the
majority of SCR stars in Figure~\ref{redpromo}.  These are presumably
K and M type subdwarfs.  A sample of 64 potential SCR red subdwarfs
from Paper XII and this paper is listed in its entirety in
Table~\ref{subdwarfs}.  These have been selected by identifying stars
in Figure~\ref{redpromo} (and the twin figure in Paper XII) with
$R_{59F}-J >$ 1.0 and having $H_R$ within 4.0 mag of the dashed line
separating the white dwarfs from the subdwarfs.  All of these have
distances estimated to be 90 pc or greater, but they are, in fact,
likely to be much closer than estimated (i.e. the distances are
incorrect), so their distance estimates are bracketed in
Table~\ref{scr-tbl} to distinguish them from presumed dwarfs whose
distance estimates are plausibly accurate.  In the same regions in
Figure~\ref{redpromo} and the complimentary figure in Paper XII, there
are 255 red subdwarf candidates from previous proper motion searches.
Thus, we have made a 25\% increase to the red subdwarf candidate pool
in the southern sky.  Future spectroscopic efforts will reveal whether
or not these candidates are true low metallicity red subdwarfs and
allow us to continue building a sizeable sample of these rare
interlopers to the solar neighborhood, very few of which have accurate
parallaxes, photometry, and spectral types.

%%%%%%%%%%%%%%%%%%%%%%%%%%%%%%%%%%%%%%%%%%%%%%%%%%%%%%%%%%%%%%%%%%%%%%%%%%%%%%
\subsection {White Dwarfs}
%%%%%%%%%%%%%%%%%%%%%%%%%%%%%%%%%%%%%%%%%%%%%%%%%%%%%%%%%%%%%%%%%%%%%%%%%%%%%%

Six of the SCR stars reported here have no distance estimates, usually
because their colors are too blue for the photometric distance
relations.  SCR 0927-4137, SCR 1811-4239, and SCR 1857-4309 are
possibly early type subdwarfs, and SCR 1800-0431B is a companion with
blended plate photometry (see $\S$ 5.5).  The remaining two stars, SCR
0818-3110 and SCR 1246-1236 are white dwarf candidates.  Both lie
solidly in the white dwarf region of Figure~\ref{redpromo}.  Two
additional SCR stars, SCR 0125-4545 and SCR 0753-2524 (a companion to
LTT 2976) also fall clearly in the white dwarf region of
Figure~\ref{redpromo} and have the largest distance estimates
presented here (516 pc and 365 pc, respectively).  These objects have
colors that are within the color ranges of the photometric distance
relations but are presumably white dwarfs with erroneous distance
estimates, hence the brackets in Table~\ref{scr-tbl}.  Two additional
objects --- SCR 1227-4541 and SCR 1916-3638 --- are near the dotted
line dividing the probable white dwarf and subdwarf regions.  We have
preliminary CCD $VRI$ photometry indicating that these two objects are
unlikely to be single white dwarfs.

Of the four white dwarf candidates, we have so far obtained
spectroscopy on only SCR 1246-1236 and confirm it to be a white dwarf.
Another candidate (SCR 0753-2524) is a companion to a star of known
distance (thereby ruling it out as a more distant early-type dwarf).
Spectroscopic observations for all white dwarf candidates are desired
for definitive confirmation.  Spectra for SCR 1246-1236 as well as for
a handful of the objects labeled with open triangles in
Figure~\ref{redpromo} as ``known'' objects that we have confirmed to
be white dwarfs for the first time will be presented in a future
publication.

Using the single color linear fit of \citet{2001Sci...292..698O} and
their error of 20\% for distances, we estimate the following distances
for the four white dwarf candidates reported in this paper: 24.7 $\pm$
4.9 pc for SCR 0125-4545, 16.2 $\pm$ 3.2 pc for SCR 0753-2524, 13.1
$\pm$ 2.6 pc for SCR 0818-3110 and 40.0 $\pm$ 8.0 pc for SCR
1246-1236.  Should these distance estimates hold true, three of the
four objects lie within the 25 pc horizon, a volume in which there are
only 109 white dwarfs currently having trigonometric parallaxes.
Trigonometric parallax observations via CTIOPI are under way to
determine accurate distances.

%%%%%%%%%%%%%%%%%%%%%%%%%%%%%%%%%%%%%%%%%%%%%%%%%%%%%%%%%%%%%%%%%%%%%%%%%%%%%%
\subsection {Comments on Individual Systems}
%%%%%%%%%%%%%%%%%%%%%%%%%%%%%%%%%%%%%%%%%%%%%%%%%%%%%%%%%%%%%%%%%%%%%%%%%%%%%%

Here we highlight systems included in this portion of the SCR survey,
most of which are multiple systems.

SCR 0640-0552 ($\mu$ $=$ 0$\farcs$592 yr$^{-1}$ at position angle
170.5$^\circ$) is the brightest new detection, with $R_{59F}$ $=$ 8.8
and an estimated distance of 8.5 pc.  CCD photometry from two nights
indicates $V_J$ $=$ 10.21, $R_{KC}$ $=$ 9.21, and $I_{KC}$ $=$ 8.03,
confirming that it is a very bright object.  These values, when
combined with the 2MASS $JHK_s$ magnitudes yield a distance estimate
of 9.4 pc using the relations in \citet{2004AJ....128.2460H}.

SCR 0753-2524 ($\mu$ $=$ 0$\farcs$426 yr$^{-1}$ at position angle
300.2$^\circ$) is a common proper motion companion to LTT 2976, which
has $\mu$ $=$ 0$\farcs$361 yr$^{-1}$ at position angle 303.7$^\circ$
and a trigonometric parallax of 0$\farcs$05116 $\pm$ 0$\farcs$00157
$=$ 19.5 $\pm$ 0.6 pc \citep{1997yCat.1239....0E}.  The separation of
the two stars is 400$\arcsec$ (projected separation $\sim$8000 AU) at
position angle 208.9$^\circ$.  SCR 0753-2524 is probably a white
dwarf, for which we estimate a distance of 16.2 pc.  This distance is
consistent with the trigonometric distance measured for LTT 2976
within the $\sim$20\% errors of the white dwarf distance relations.
Although the sizes of the proper motions do not match perfectly, the
better determined position angles are consistent, so we conclude that
the two stars form a system.

SCR 1510-4259 ($\mu$ $=$ 0$\farcs$430 yr$^{-1}$ at position angle
229.0$^\circ$) is a common proper motion companion to CD $-$42 10084
($\mu$ $=$ 0$\farcs$436 yr$^{-1}$ at position angle 228.1$^\circ$),
for which Hipparcos measured a trigonometric parallax of 0.03999 $\pm$
0.00241 $=$ 25.0 $\pm$ 1.5 pc \citep{1997yCat.1239....0E}.  The
separation of the two stars is 88$\arcsec$ at position angle
123.5$^\circ$.  The distance estimate for SCR 1510-4259, 31.2 pc, is
consistent with the Hipparcos distance for CD $-$42 10084, within the
26\% errors of the plate magnitude distance relations, and the proper
motions are a match.  We conclude that the two stars form a system.

SCR 1529-4238 ($\mu$ $=$ 0$\farcs$447 yr$^{-1}$ at position angle
243.2$^\circ$) is a probable common proper motion companion to L
408-087 ($\mu$ $=$ 0$\farcs$285 yr$^{-1}$ at position angle
235.0$^\circ$) (NLTT Catalog) for which there is no trigonometric
parallax available.  The separation of the two stars is 45$\arcsec$ at
position angle 159.0$^\circ$.  The sizes of the proper motions do not
match well, but the position angles are a fair match.  Given the
incomplete information in the NLTT (no photographic R magnitude),
presumably because of the very crowded field, the proper motion for L
408-087 is suspect.  In fact, we cannot estimate a distance for L
408-087 because it is blended on several plates, precluding reliable
plate magnitudes.  We tentatively conclude that the two stars form a
system.

SCR 1608-2913 AB ($\mu$ $=$ 0$\farcs$540 yr$^{-1}$ at position angle
231.0$^\circ$) is a close double system with separation 2$\farcs$5 at
position angle 266.2$^\circ$, determined using frames acquired during
CTIOPI.  The magnitude differences are 0.56, 0.49, and 0.37 mag at
$VRI$, respectively.

SCR 1800-0431 AB ($\mu$ $=$ 0$\farcs$402 yr$^{-1}$ at position angle
227.4$^\circ$) is a common proper motion pair with a separation of
24$\arcsec$ at position angle 234.0$^\circ$.  While investigating the
primary, the B component was noticed on images extracted from all four
available plates; however, it is blended with other sources in all
four cases so no reliable plate photometry or distance estimate is
available.

SCR1856-1951 is one of nine objects with a distance estimate in excess
of 200 pc and the only one not flagged as a subdwarf candidate.  The
$R_{59F}-J$ color is too blue for the conservative subdwarf candidate
selection criteria.  While this object may be a subdwarf, dwarf
contamination at these colors warrants exclusion from the subdwarf
candidate list.

SCR 2123-3653 ($\mu$ $=$ 0$\farcs$446 yr$^{-1}$ at position angle
133.7$^\circ$) is a common proper motion companion to LTT 8495 ($\mu$
$=$ 0$\farcs$417 yr$^{-1}$ at position angle 134.1$^\circ$), for which
there is no trigonometric parallax available.  The separation of the
two stars is 50$\arcsec$ at position angle 168.0$^\circ$.  The proper
motions are consistent, indicating that the two stars almost certainly
form a system.  However, from plate$+$$JHK_s$ photometry, distance
estimates are 25.9 pc and 78.5 pc for LTT 8495 and SCR 2123-3653,
respectively, which indicates that if the two are a pair, LTT 8495 is
likely to be an unresolved multiple.

%%%%%%%%%%%%%%%%%%%%%%%%%%%%%%%%%%%%%%%%%%%%%%%%%%%%%%%%%%%%%%%%%%%%%%%%%%%%%%
\section {Discussion}
%%%%%%%%%%%%%%%%%%%%%%%%%%%%%%%%%%%%%%%%%%%%%%%%%%%%%%%%%%%%%%%%%%%%%%%%%%%%%%

One of the primary motivations for high proper motion surveys is, of
course, the promise of detecting new nearby stars.  The new nearby
discoveries are typically red dwarfs, and occasionally, white dwarfs.
The output lists of sources detected, once culled for false hits, also
includes subdwarfs of very high intrinsic velocity that are generally
not as near as their main sequence counterparts, but are nevertheless
interesting in their own right as tracers of the Galactic halo
population.

Listed in Table~\ref{diststats} is a summary of the number of SCR
systems with distance estimates within each of our two target horizons
(10 pc and 25 pc) and beyond.  New common proper motion objects that
are companions to known objects are not included in the counts, nor
are probable white dwarfs (because their distance estimates require
different relations than applied here).  The two numbers given for
each entry represent the number of SCR systems reported in Paper XII
and this paper, respectively.  In most cases the numbers are
comparable, which reflects the fact that although much of the sky
searched in this paper has already been searched by Giclas and Luyten,
there is significantly more sky covered in this paper than in Paper
XII.

In total, we have found 43 new candidate systems within 25 pc of the
Sun.  There remain several likely subdwarfs with overestimated
distances that may fall in closer bins than indicated in
Table~\ref{diststats}.  Perhaps the most surprising result of this
survey is the discovery that the slowest proper motion (0$\farcs$6
yr$^{-1}$ $>$ $\mu$ $\ge$ 0$\farcs$4 yr$^{-1}$) contains the largest
number (26) of new candidates for systems within 25 pc.  In fact, we
have found equal numbers of 10 pc candidates with $\mu$ $>$ 1$\farcs$0
yr$^{-1}$ as we have with 0$\farcs$6 yr$^{-1}$ $>$ $\mu$ $\ge$
0$\farcs$4 yr$^{-1}$.  As pointed out in Paper XII, the presence of so
many new nearby stars with relatively low proper motions hints that
there may be large numbers of even slower moving stars that remain
hidden in the solar neighborhood.  Thus, searches for nearby stars
buried in large samples with smaller proper motions are warranted, in
particular given the availability of large photometric databases that
allow the derivation of accurate distance estimates when optical and
infrared data are combined, such as done here.

In summary, we have revealed a total of 299 new SCR proper motion
systems in the southern sky.  Of these, 148 have $\mu$ $\ge$
0$\farcs$5 yr$^{-1}$, making them new members of the classic LHS
sample.  Among the new discoveries, we anticipate that most are main
sequence M dwarfs, at least nine are white dwarf candidates, at least
five are new binary systems, and 64 are K or M type subdwarf
candidates.  Seven additional proper motion companions to previously
known HPM stars were also found.  Five of the nine white dwarf
candidates are anticipated to be within 25 pc.  Worthy of note are the
eight new SCR stars brighter than $R_{59F}$ $=$ 12, six of which have
$\mu$ $\ge$ 0$\farcs$5 yr$^{-1}$, hinting at the possibility of
relatively bright nearby stars that have not yet been identified.

All three sets of stars --- white dwarfs, red dwarfs, and subdwarfs
--- provide important contributions to these intrinsically faint,
neglected samples.  Undoubtedly, objects fainter than our survey
cutoff of $R_{59F}$ $=$ 16.5 remain to be found, as well as a small
number of stars meeting our survey criteria that fell in crowded
regions or were simply missed because of the stringent limits required
for SCR star veracity.  Finally, we are delighted to have discovered
during the SCR survey five new systems that are likely new members of
the RECONS 10 pc sample, and are actively determining accurate
parallaxes for them, as well as for many of the 38 other SCR systems
within 25 pc, via our parallax program in Chile, CTIOPI.

%%%%%%%%%%%%%%%%%%%%%%%%%%%%%%%%%%%%%%%%%%%%%%%%%%%%%%%%%%%%%%%%%%%%%%%%%%%%%%
\section {Acknowledgments}
%%%%%%%%%%%%%%%%%%%%%%%%%%%%%%%%%%%%%%%%%%%%%%%%%%%%%%%%%%%%%%%%%%%%%%%%%%%%%%

Funding for the SuperCOSMOS Sky Survey is provided by the UK Particle
Physics and Astronomy Research Council.  N.C.H. would like to thank
colleagues in the Wide Field Astronomy Unit at Edinburgh for their
work in making the SSS possible; particular thanks go to Mike Read,
Sue Tritton, and Harvey MacGillivray.  The authors would like to thank
the referee for constructive comments aiding to the clarity of this
paper.  The RECONS team at Georgia State University wishes to thank
NASA's Space Interferometry Mission for its continued support of our
study of nearby stars.  This work has made use of the SIMBAD, VizieR,
and Aladin databases, operated at the CDS in Strasbourg, France.  We
have also used data products from the Two Micron All Sky Survey, which
is a joint project of the University of Massachusetts and the Infrared
Processing and Analysis Center, funded by NASA and NSF.

%%%%%%%%%%%%%%%%%%%%%%%%%%%%%%%%%%%% REFS %%%%%%%%%%%%%%%%%%%%%%%%%%%%%%%%%%%%

\clearpage

%%%%%%%%%%%%%%%%%%%%%%%%%%% FIGURE CAPTIONS %%%%%%%%%%%%%%%%%%%%%%%%%%%%%%%%%%

\figcaption[subasavage.fig01.ps]{Plate coverage of the entire SCR
survey, including sky coverage in Paper XII and this paper.  Plates
colored in white were excluded from the search, primarily because of
source crowding (Galactic plane, LMC, SMC) or a limited time span
between plates.
\label{pltfields}}

\figcaption[subasavage.fig02.ps]{Sky distribution of new LHS objects
from recent proper motion surveys.  Only stars with $\mu$ $\ge$
0$\farcs$5 yr$^{-1}$ are plotted.  Filled triangles are from the SCR survey
discussed here and in Paper XII.  An 'X' represents an object from the
SUPERBLINK survey.  Other symbols represent objects from other
surveys: WT (open circles), Scholz (open triangles), Calan-ESO (open
boxes), Oppenheimer (open stars), SIPS (open diamonds).  The curve
represents the Galactic plane, where more high proper motion objects
are yet to be revealed.
\label{surveys}}

\figcaption[subasavage.fig03.ps]{Color-apparent magnitude diagram for
the SCR systems with $\mu$ $\ge$ 0$\farcs$4 yr$^{-1}$ (size of the
points splits SCR sample into stars with $\mu$ more or less than
0$\farcs$5 yr$^{-1}$) and known systems with $\mu$ $\ge$ 0$\farcs$5
yr$^{-1}$ found in this portion of the SCR survey, declinations
$-$47$^\circ$ to 00$^\circ$.  Triangles indicate confirmed white
dwarfs.
\label{colmag}}

\figcaption[subasavage.fig04.ps]{Reduced proper motion diagram for the
SCR systems with $\mu$ $\ge$ 0$\farcs$4 yr$^{-1}$ (size of the points
splits SCR sample into stars with $\mu$ more or less than 0$\farcs$5
yr$^{-1}$) and known systems with $\mu$ $\ge$ 0$\farcs$5 yr$^{-1}$
found in this portion of the SCR survey, declinations $-$47$^\circ$ to
00$^\circ$.  Reduced proper motion (vertical axis) has units
of magnitudes.  The dashed line serves merely as a reference to
distinguish white dwarfs from subdwarfs.  Triangles indicate confirmed
white dwarfs.
\label{redpromo}}

\figcaption[subasavage.fig05.ps]{Finder charts for the 152 new SCR
systems and four wide companions to known HPM stars.  Each finder is
5$\arcmin$ on a side, with north up and east to the left.  The
observation epoch for each frame is given.  The bandpass for each
finder is $R_{59F}$.
\label{finders}}

%%%%%%%%%%%%%%%%%%%%%%%%%%%%%%%%%%% FIGURES %%%%%%%%%%%%%%%%%%%%%%%%%%%%%%%%%%

\begin{figure}
\plotone{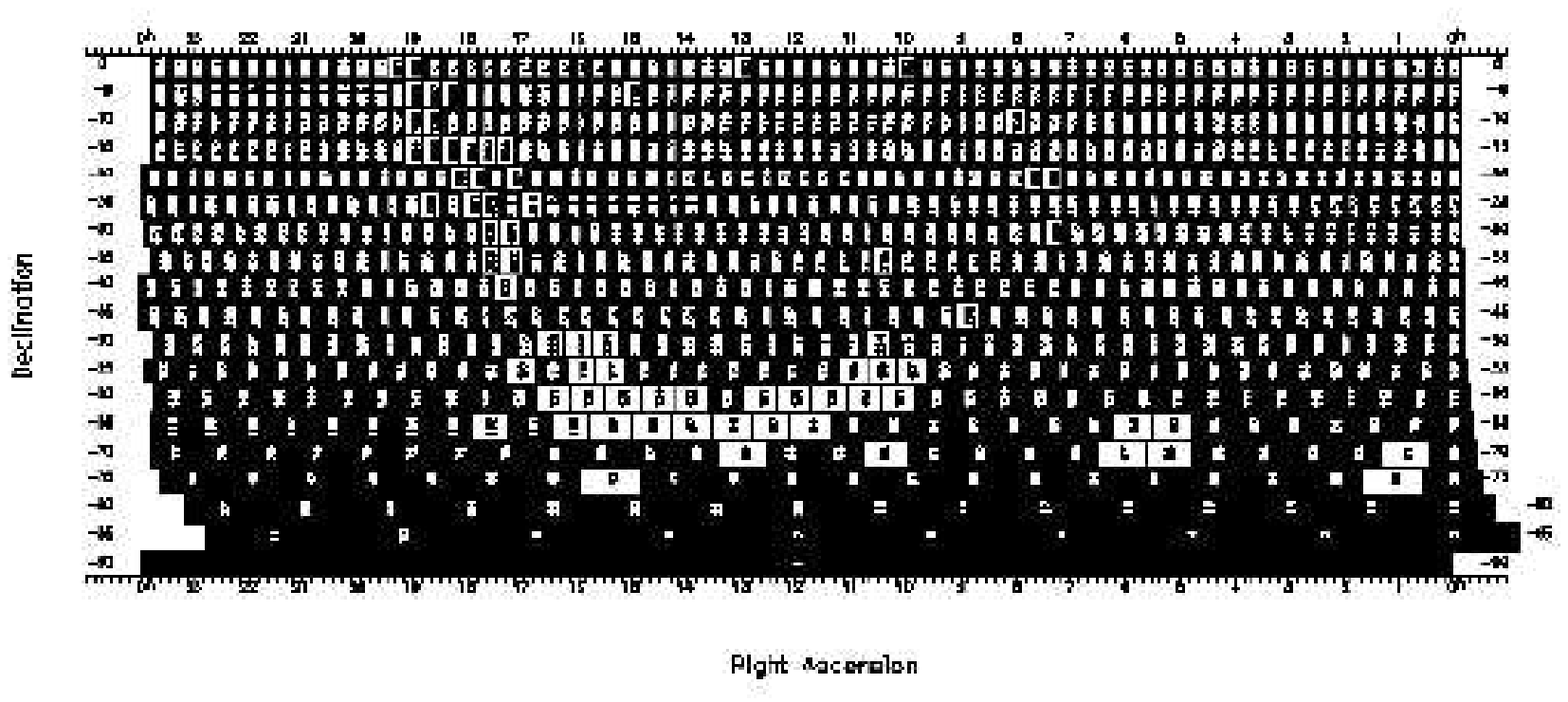}
%\label{pltfields}
\end{figure}

\clearpage

\begin{figure}
\plotone{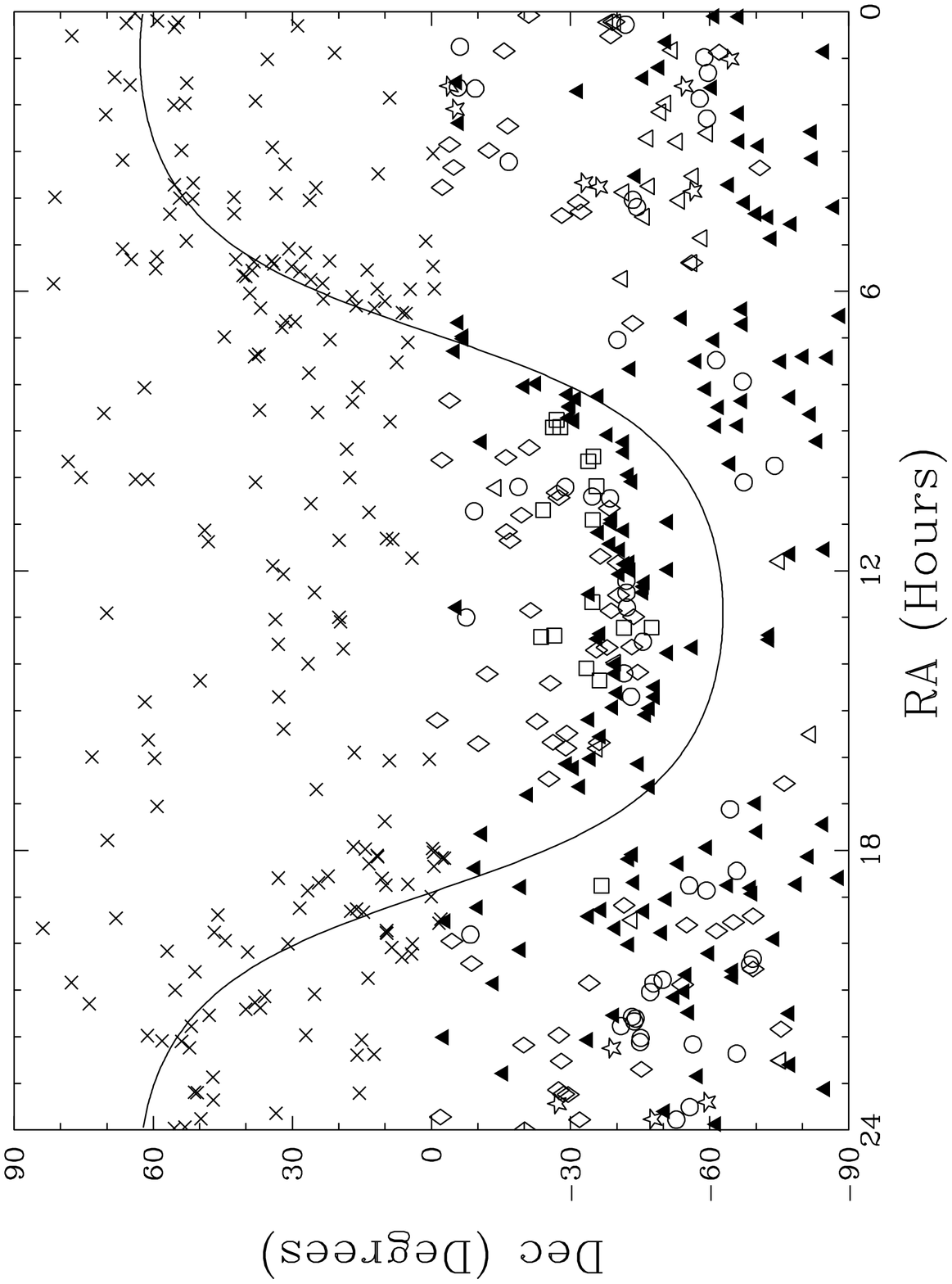}
%\label{surveys}
\end{figure}

\clearpage

\begin{figure}
\plotone{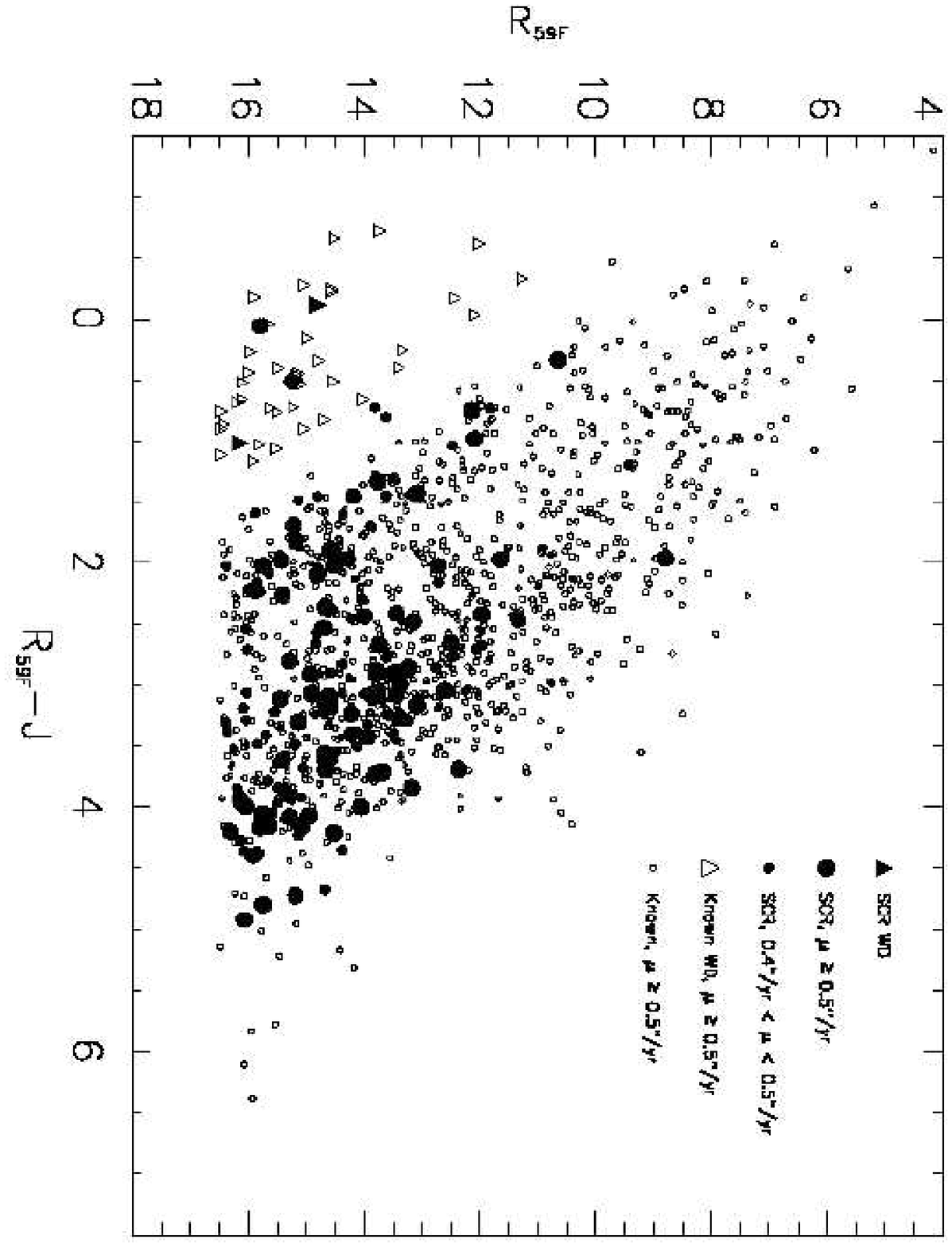}
%\label{colmag}
\end{figure}

\clearpage

\begin{figure}
\plotone{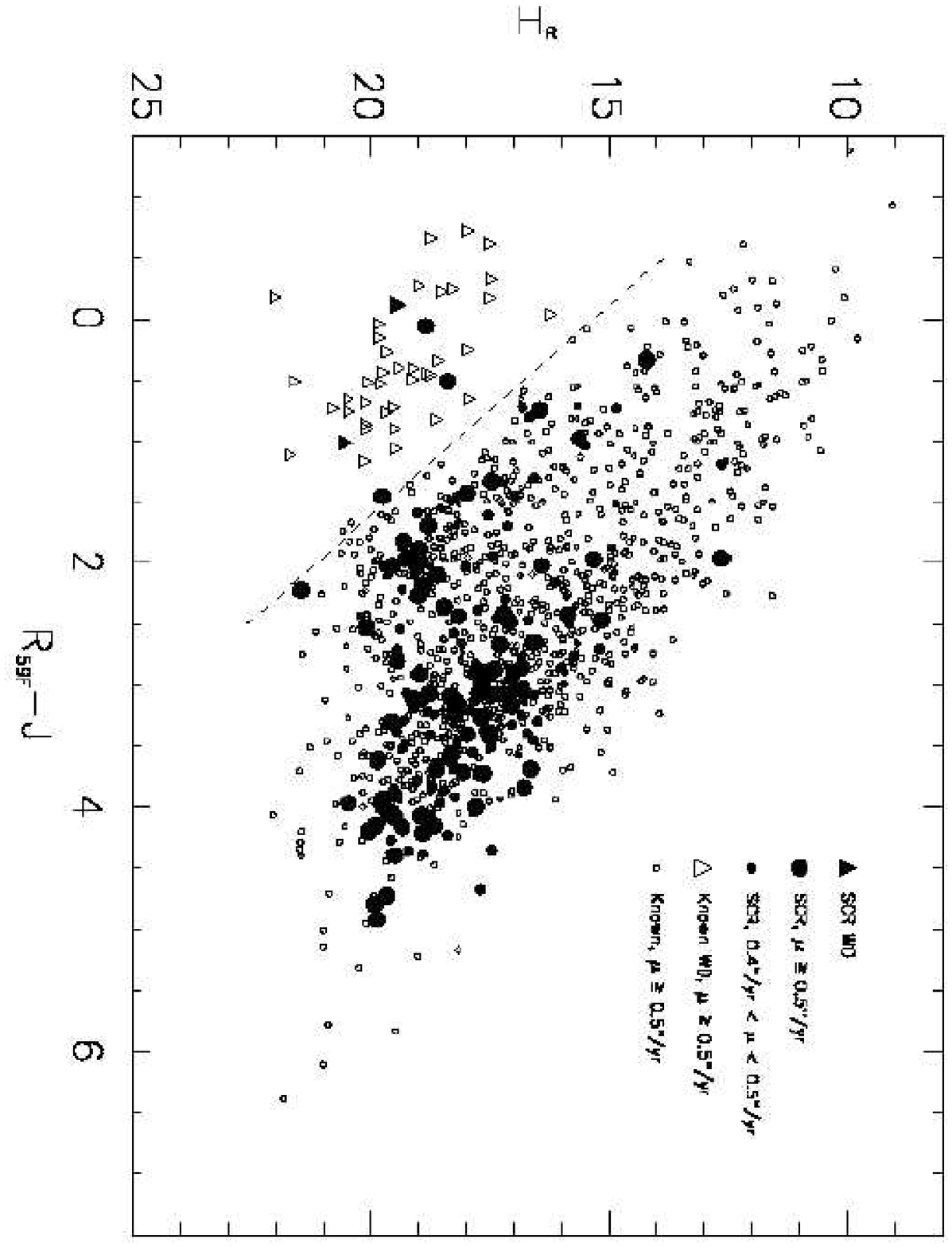}
%\label{redpromo}
\end{figure}

\clearpage

\begin{figure}
\plotone{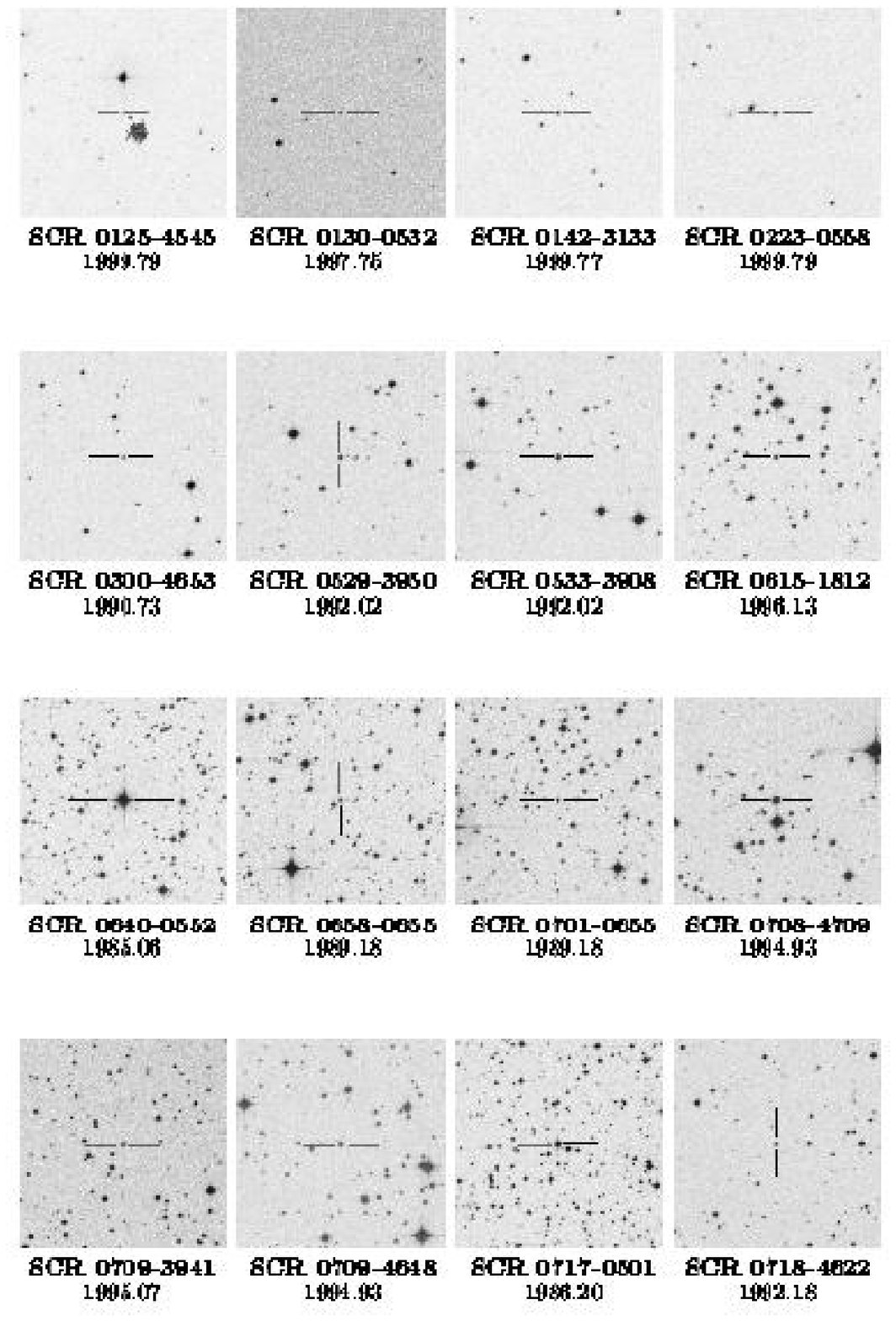}
%\label{finders}
\end{figure}

\clearpage

\begin{figure}
\plotone{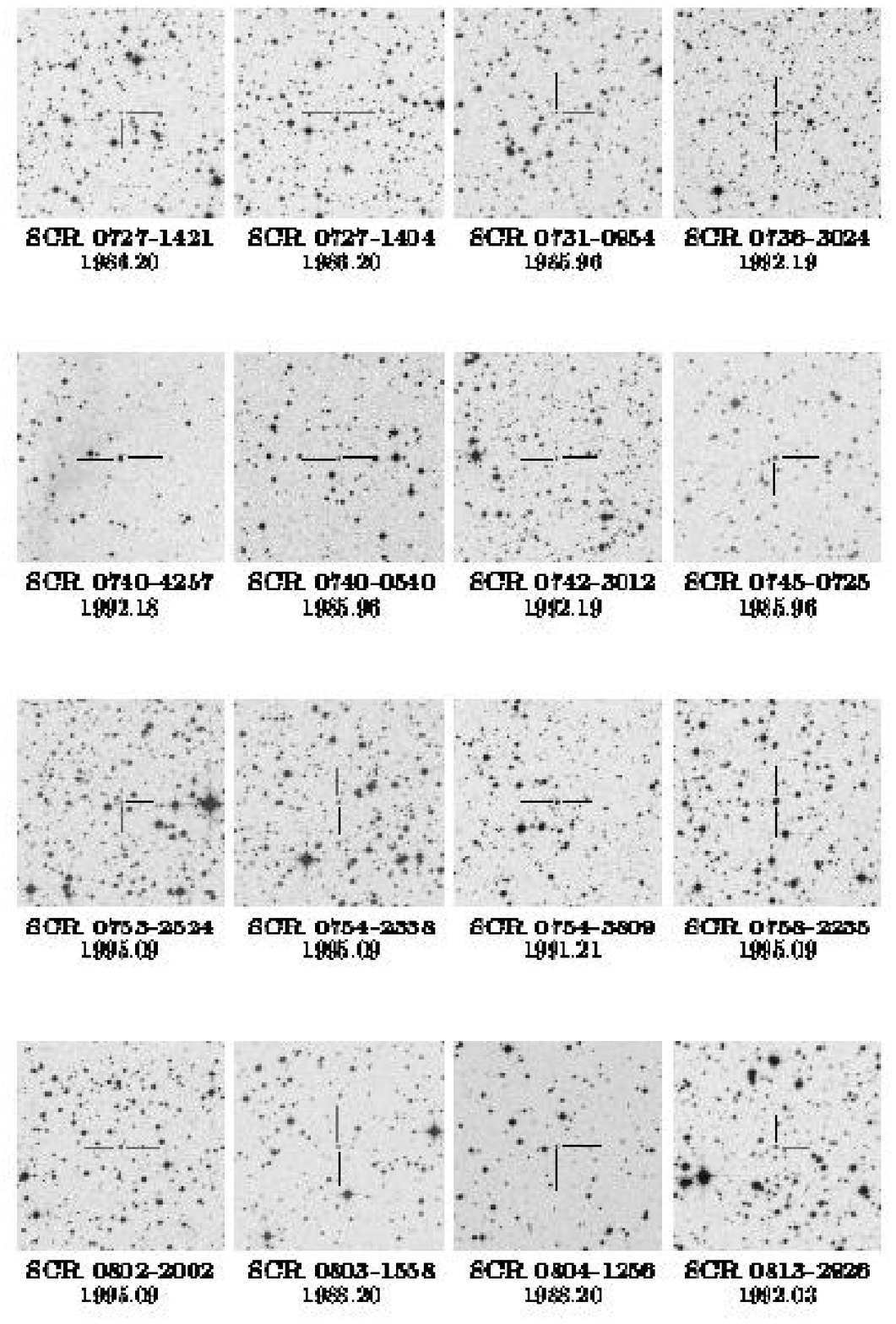}
%\label{finders}
\end{figure}

\clearpage

\begin{figure}
\plotone{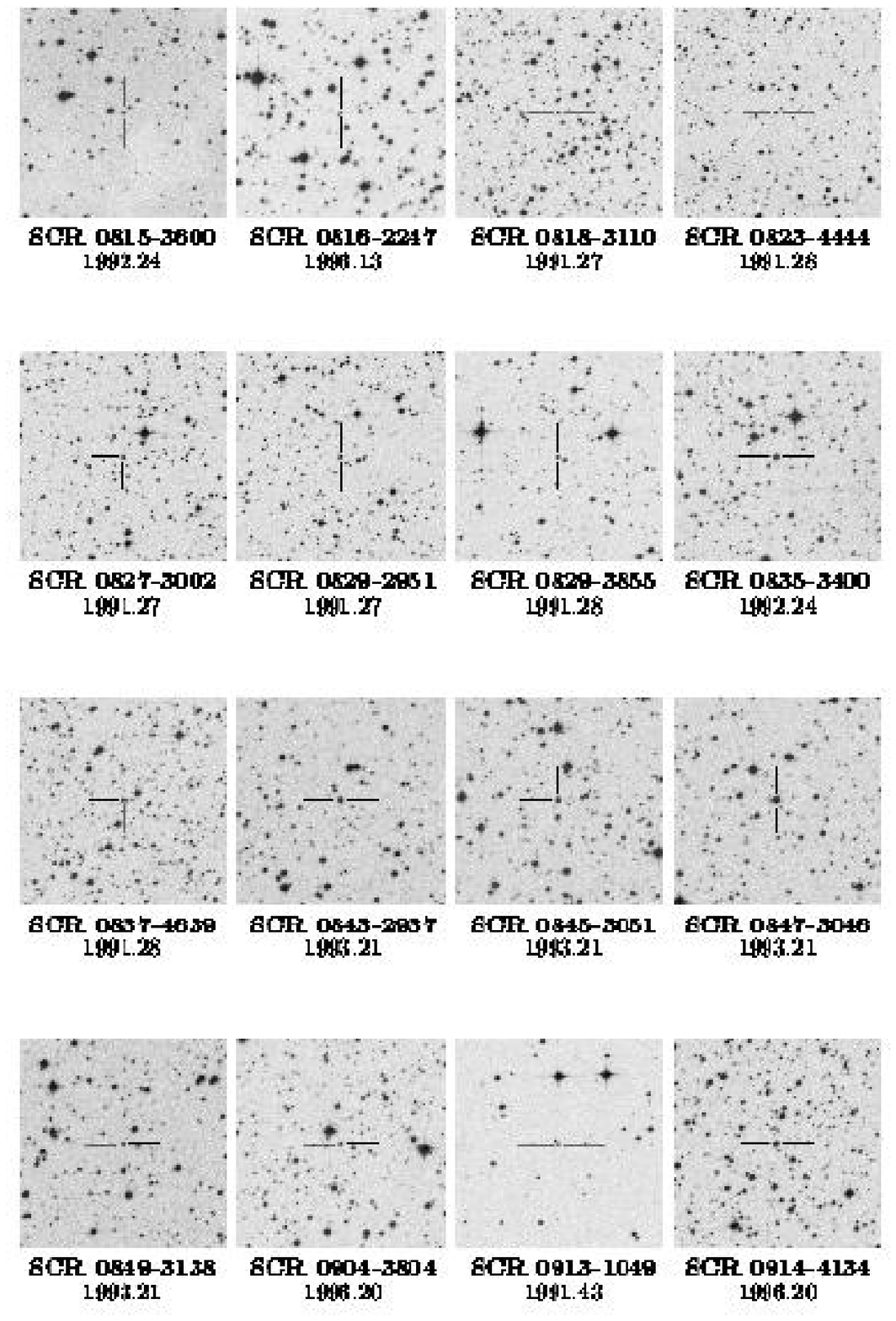}
%\label{finders}
\end{figure}

\clearpage

\begin{figure}
\plotone{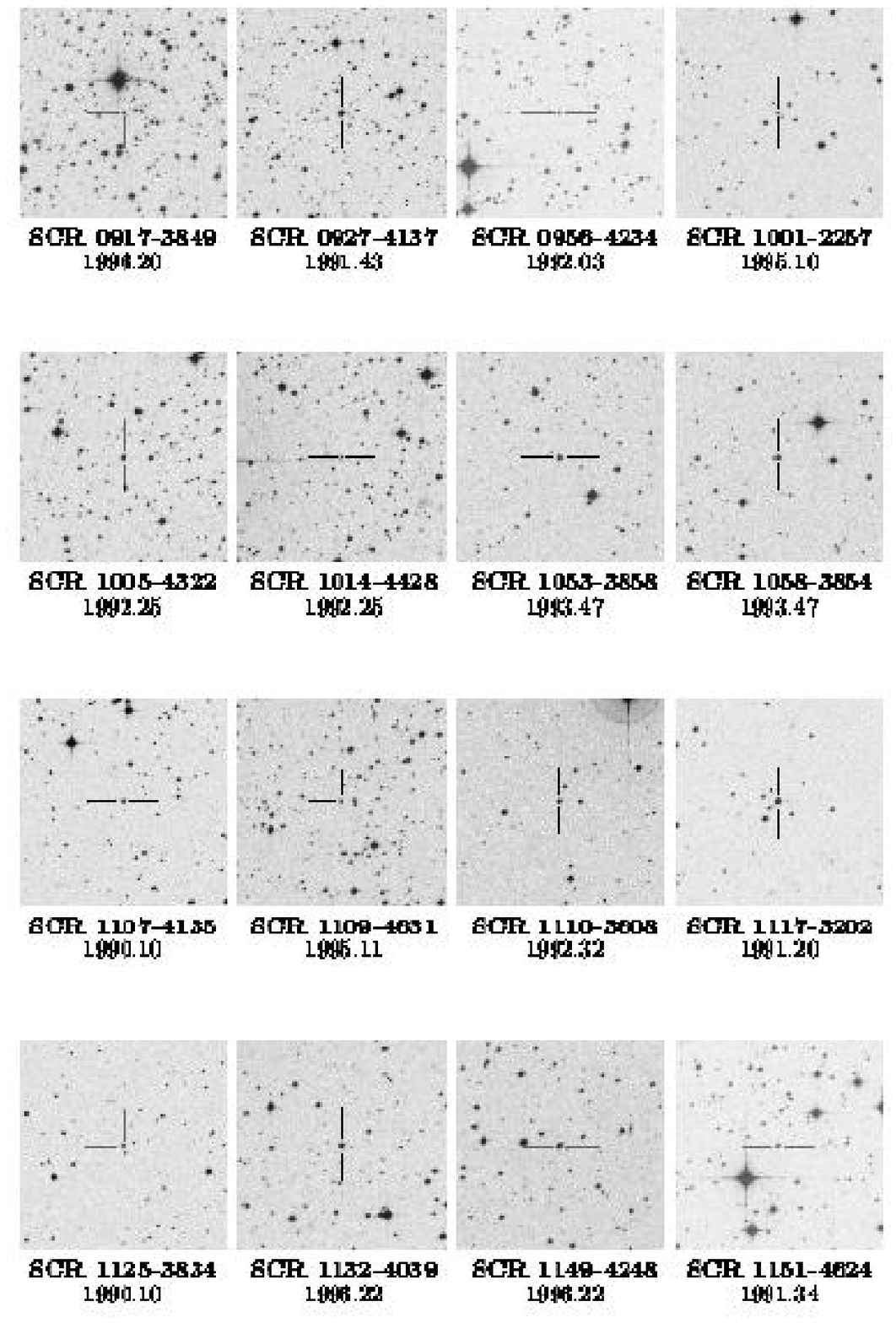}
%\label{finders}
\end{figure}

\clearpage

\begin{figure}
\plotone{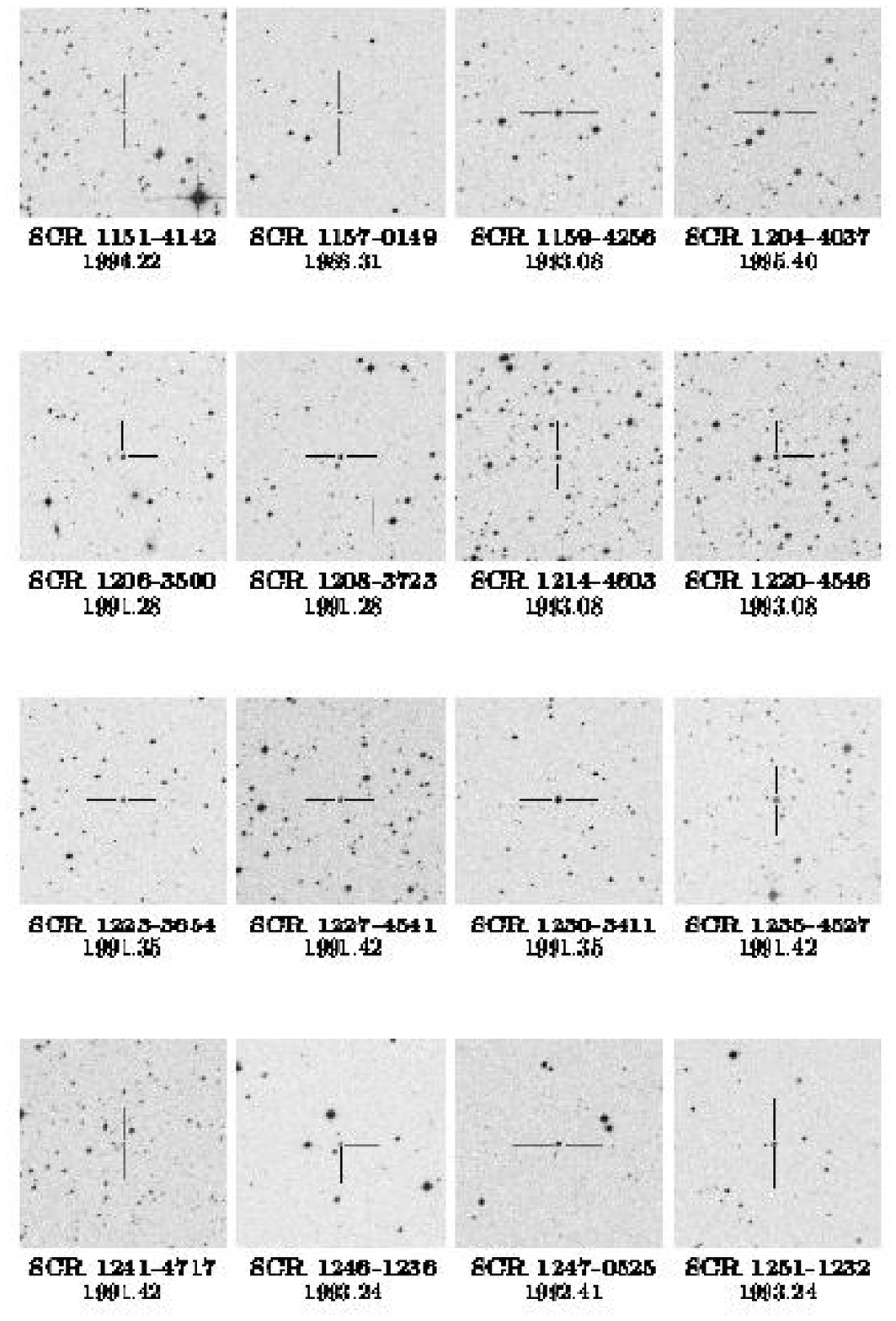}
%\label{finders}
\end{figure}

\clearpage

\begin{figure}
\plotone{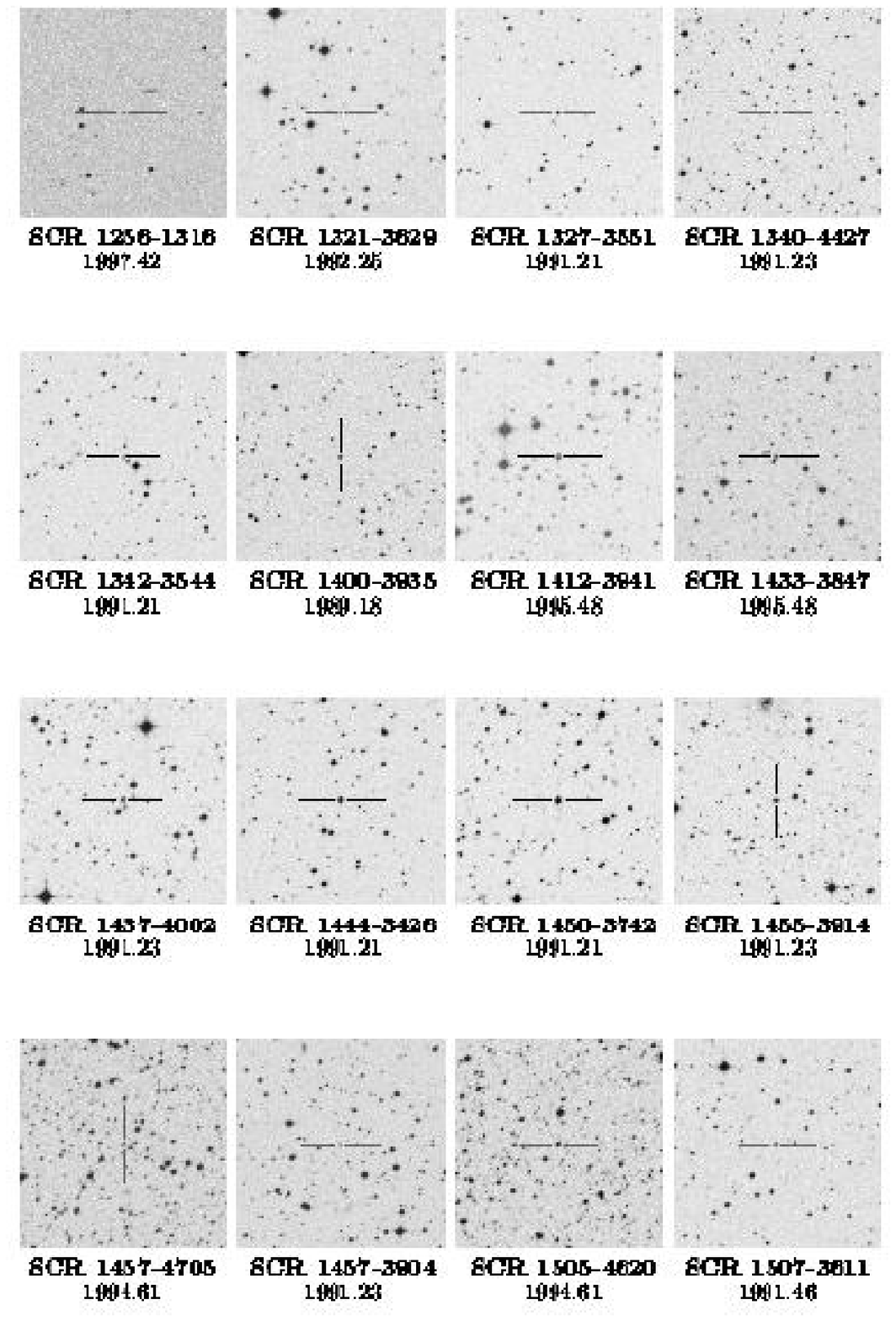}
%\label{finders}
\end{figure}

\clearpage

\begin{figure}
\plotone{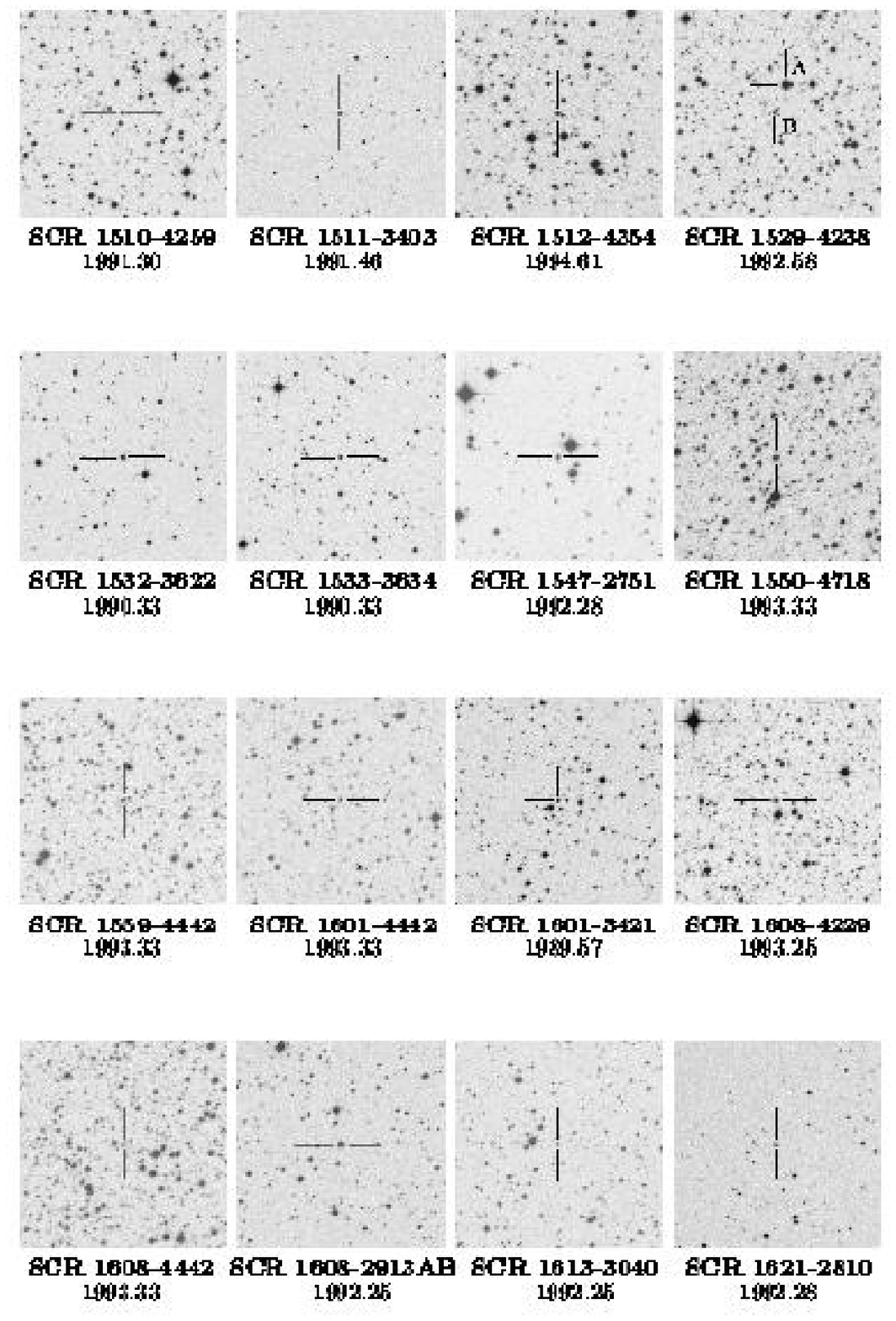}
%\label{finders}
\end{figure}

\clearpage

\begin{figure}
\plotone{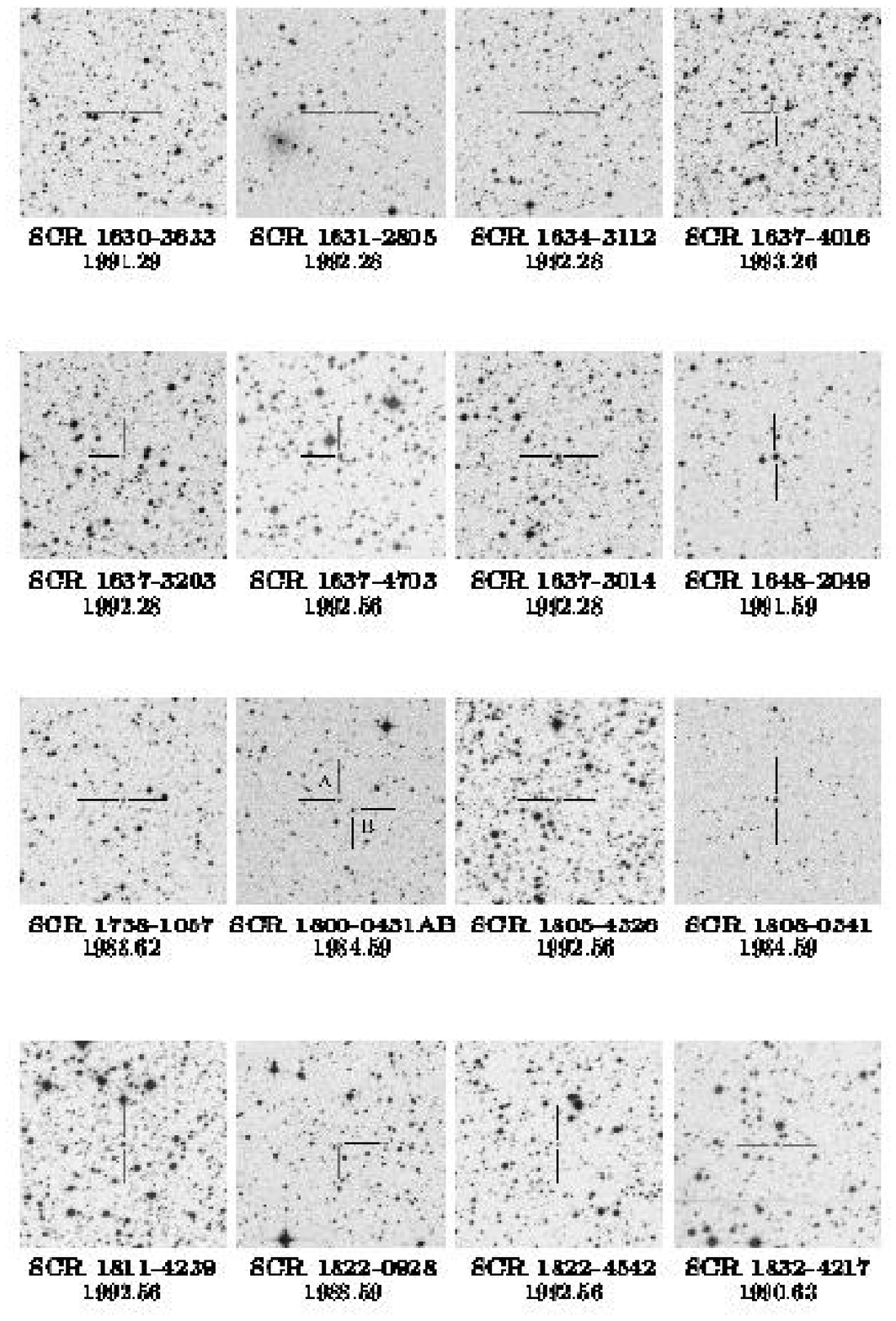}
%\label{finders}
\end{figure}

\clearpage

\begin{figure}
\plotone{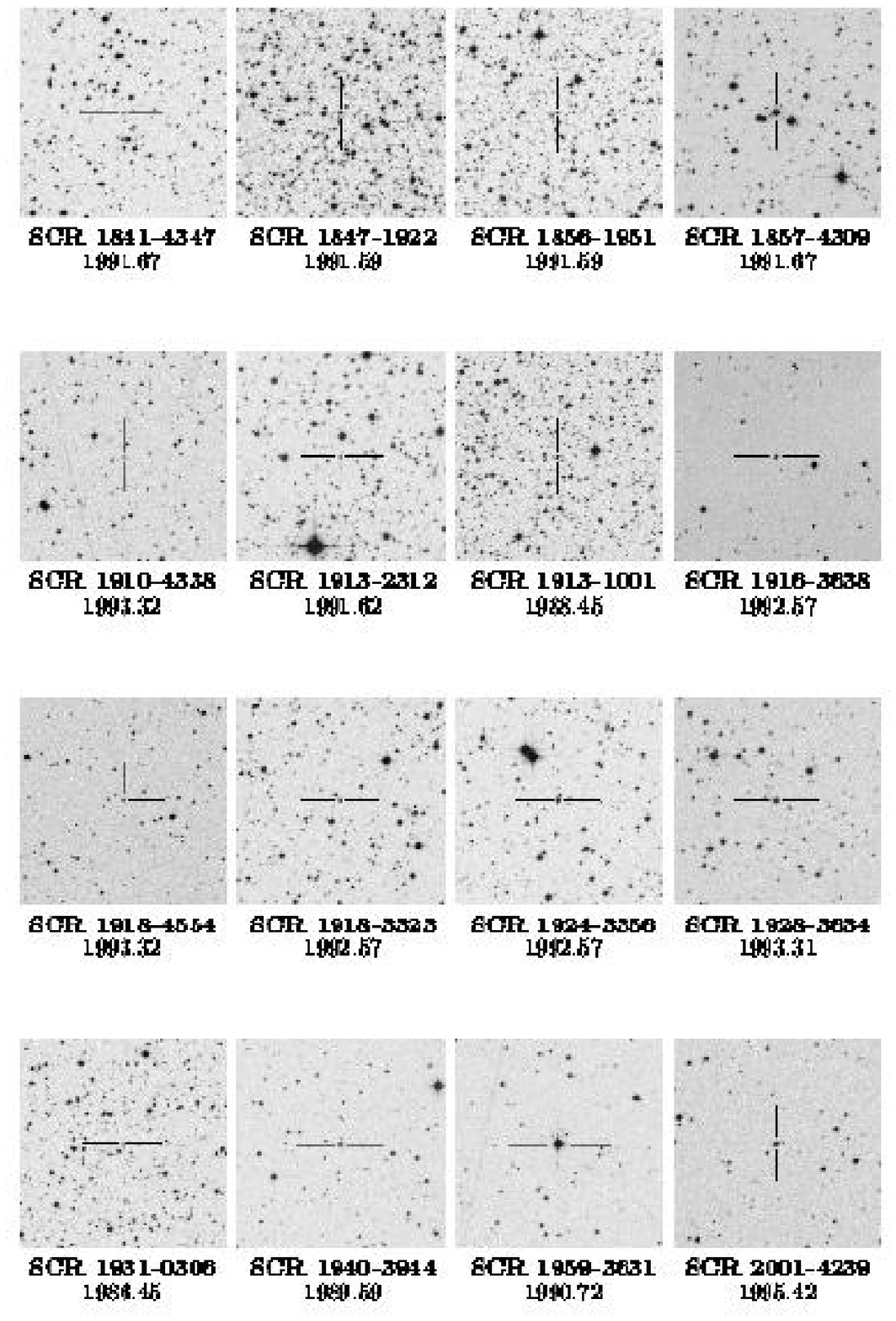}
%\label{finders}
\end{figure}

\clearpage

\begin{figure}
\plotone{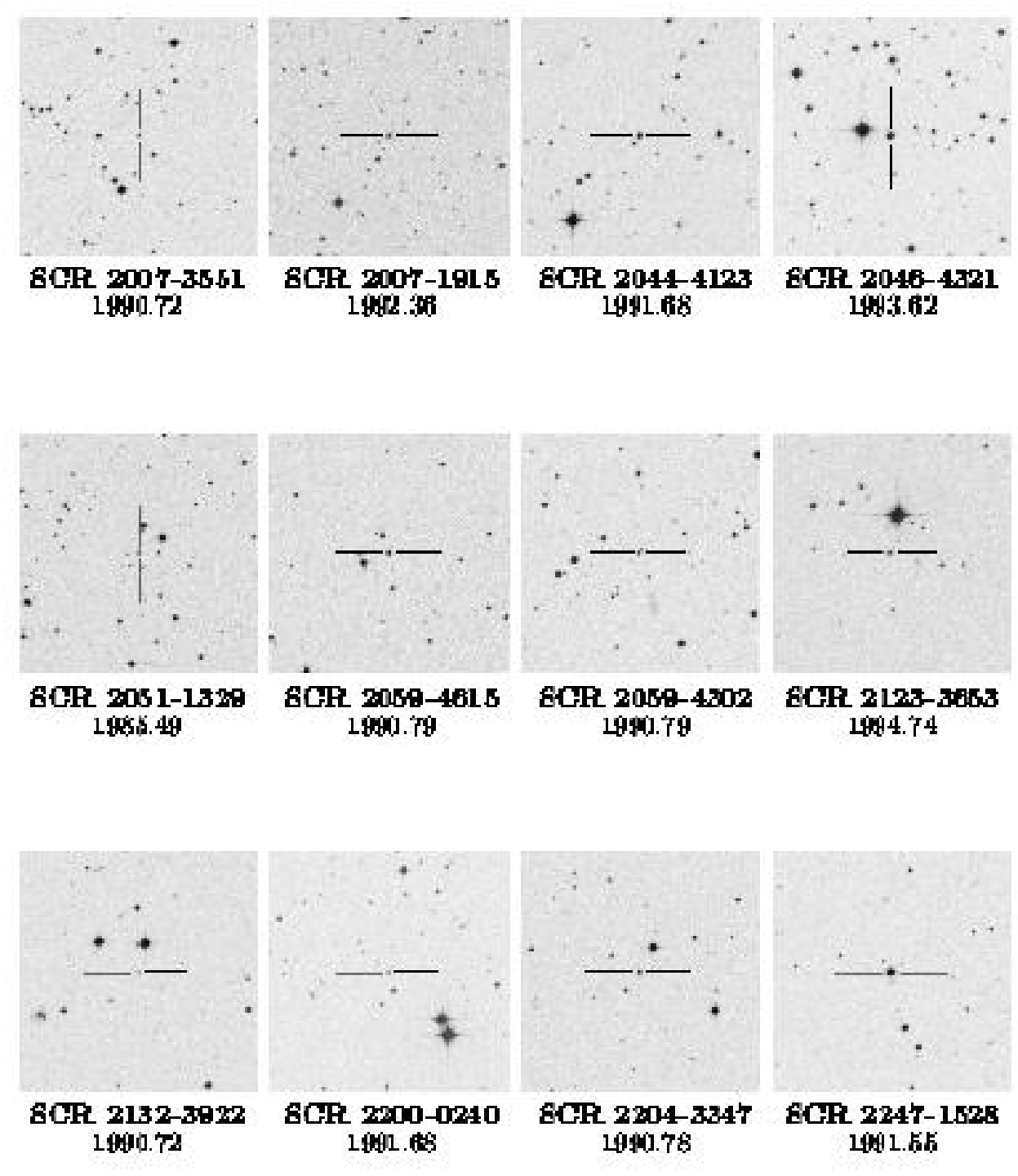}
%\label{finders}
\end{figure}

%%%%%%%%%%%%%%%%%%%%%%%%%%%%%% TABLE1: NEW LHS MEMBERS %%%%%%%%%%%%%%%%%%%%%%%%%%

%%\voffset100pt{}
\begin{deluxetable}{lcccc}
\tabletypesize{\footnotesize}
\tablecaption{Proper Motion Surveys and Number of New LHS Objects Discovered. 
\label{pm-surveys}}
\tablewidth{0pt}

\tablehead{%%\vspace{-25pt} \\
           \colhead{Survey}&
           \colhead{$\mu$ $\ge$ 1$\farcs$0 yr$^{-1}$}&
           \colhead{1$\farcs$0 yr$^{-1}$ $>$ $\mu$ $\ge$ 0$\farcs$5 yr$^{-1}$}&
           \colhead{Total}&
           \colhead{\# of Publications \tablenotemark{a}}}

\startdata
LHS                                &     528 &    3074 &    3602 &   1  \\
SUPERBLINK                         &      18 &     180 &     198 &   2  \\
SuperCOSMOS-RECONS                 &       9 &     141 &     150 &   4  \\
SIPS (Deacon et al.)               &      10 &      58 &      68 &   1  \\
WT (Wroblewski and collaborators)  &       2 &      46 &      48 &   7  \\
Scholz and collaborators           &       5 &      21 &      26 &   3  \\
Calan-ESO (Ruiz and collaborators) &       3 &      14 &      17 &   2  \\
Oppenheimer et al.                 &       3 &       8 &      11 &   1  \\
Pokorny et al.                     & unknown & unknown & unknown &   2  \\

\enddata

\tablenotetext{a}{references include 
\citet{1979lccs.book.....L},
\citet{2002AJ....124.1190L,2003AJ....126..921L},
\citet{2004AJ....128..437H,2004AJ....128.2460H,2005AJ....129..413S}, this paper,
\citet{2005A&A...435..363D},
\citet{1989A&AS...78..231W,1991A&AS...91..129W,1994A&AS..105..179W,1996A&AS..115..481W,1997A&AS..122..447W,1999A&AS..139...25W,2001A&A...367..725W},
\citet{2000A&A...353..958S,2002ApJ...565..539S,2004A&A...425..519S},
\citet{1987RMxAA..14..381R,2001ApJS..133..119R},
\citet{2001Sci...292..698O}, 
and \citet{2003A&A...397..575P,2004A&A...421..763P}}

\end{deluxetable}

\clearpage

%%%%%%%%%%%%%%%%%%%%%%%%%%%%%%%%%%%%%%%%%%%%%%%%%%%%%%%%%%%%%%%%%%%%%%%%%%%%%%
%%%%%%%%%%%%%%%%%%%%%%%%%%%% TABLE2: SCR DISCOVERIES %%%%%%%%%%%%%%%%%%%%%%%%%
%%%%%%%%%%%%%%%%%%%%%%%%%%%%%%%%%%%%%%%%%%%%%%%%%%%%%%%%%%%%%%%%%%%%%%%%%%%%%%

%%\voffset30pt{}
\begin{deluxetable}{lcccrrrrrrrcrl}

\rotate 
\tabletypesize{\tiny} 
\tablecaption{Proper motions,
photographic and infrared photometry, and distance estimates for the
SuperCOSMOS$-$RECONS sample with $\mu$ $\ge$ 0$\farcs$4 yr$^{-1}$ 
from $-$47$^\circ$ $>$ $\delta$ $>$ 00$^\circ$.
\label{scr-tbl}}
\tablewidth{0pt}

\tablehead{%%\vspace{-25pt} \\
           \colhead{Name}&
           \colhead{RA \hskip25pt DEC}&
           \colhead{$\mu$}&
	   \colhead{$\sigma_\mu$}&
	   \colhead{$\theta$}&
           \colhead{$B_J$}&
           \colhead{$R_{59F}$}&
           \colhead{$I_{IVN}$}&
           \colhead{$J$}&
           \colhead{$H$}&
           \colhead{$K_s$}&
           \colhead{$R_{59F}$ $-$ $J$}&
           \colhead{Est Dist}&
           \colhead{Notes}\\

           \colhead{}&
           \colhead{(J2000)}&
           \colhead{($\arcsec$)}&
           \colhead{($\arcsec$)}&
	   \colhead{($^\circ$)}&
           \colhead{}&
           \colhead{}&
           \colhead{}&
           \colhead{}&
           \colhead{}&
           \colhead{}&
           \colhead{}&
           \colhead{(pc)}&
           \colhead{}}

\startdata
%%\vspace{-25.05pt} \\
\tableline %%\vspace{-15pt} \\
\multicolumn{14}{c}{SuperCOSMOS$-$RECONS sample with $\mu$ $\ge$ 0$\farcs$5 yr$^{-1}$} \\
\tableline %%\vspace{-15pt} \\

  SCR 0125-4545  &  01 25 18.04 $-$45 45 31.2  &  0.759  &  0.007  &  137.8  &  17.04  &  16.13  &  15.80  &  15.11  &  14.84  &  14.91  &   1.02  & [515.7] &  Probable white dwarf \tablenotemark{a} \\
  SCR 0130-0532  &  01 30 43.82 $-$05 32 22.1  &  0.552  &  0.006  &  118.2  & \nodata &  16.05  &  13.63  &  12.06  &  11.48  &  11.19  &   3.99  &   42.4~ &  \\
  SCR 0142-3133  &  01 42 20.39 $-$31 33 35.9  &  0.749  &  0.012  &  155.1  &  18.21  &  16.11  &  13.81  &  12.15  &  11.65  &  11.36  &   3.96  &   47.2~ &  \\
  SCR 0223-0558  &  02 23 26.64 $-$05 58 47.4  &  0.530  &  0.006  &   84.5  &  17.70  &  15.47  &  13.49  &  12.36  &  11.79  &  11.56  &   3.11  &   73.4~ &  \\
  SCR 0300-4653  &  03 00 45.22 $-$46 53 50.1  &  0.779  &  0.008  &   68.7  &  17.62  &  15.40  &  12.80  &  11.79  &  11.31  &  11.02  &   3.61  &   49.5~ &  \\
  SCR 0640-0552  &  06 40 13.97 $-$05 52 23.5  &  0.592  &  0.008  &  170.5  &  11.23  &   8.79  &   7.59  &   6.84  &   6.21  &   5.96  &   1.95  &    8.5~ &  \\
  SCR 0658-0655  &  06 58 14.14 $-$06 55 35.4  &  0.574  &  0.003  &  130.6  &  16.73  &  14.68  &  12.93  &  12.33  &  11.76  &  11.53  &   2.35  & [104.9] &  \tablenotemark{a} \\
  SCR 0701-0655  &  07 01 17.79 $-$06 55 49.4  &  0.582  &  0.003  &  183.8  &  17.68  &  15.75  &  14.57  &  13.73  &  13.19  &  13.00  &   2.02  & [234.3] &  \tablenotemark{a} \\
  SCR 0717-0501  &  07 17 17.10 $-$05 01 04.0  &  0.580  &  0.004  &  133.6  &  13.86  &  11.34  &   8.83  &   8.87  &   8.35  &   8.05  &   2.47  &   15.9~ &  \\
  SCR 0740-4257  &  07 40 11.80 $-$42 57 40.1  &  0.714  &  0.013  &  318.1  &  14.52  &  12.37  &   9.99  &   8.68  &   8.09  &   7.77  &   3.69  &   10.0~ &  \\
  SCR 0758-2235  &  07 58 53.17 $-$22 35 52.8  &  0.547  &  0.012  &  153.8  &  14.87  &  12.72  &  10.81  &  10.71  &  10.19  &   9.98  &   2.01  &   56.1~ &  \\
  SCR 0802-2002  &  08 02 37.92 $-$20 02 26.4  &  0.670  &  0.008  &  139.9  &  15.81  &  13.29  &  10.61  &  10.43  &   9.80  &   9.57  &   2.86  &   32.3~ &  \\
  SCR 0813-2926  &  08 13 07.54 $-$29 26 06.9  &  0.521  &  0.006  &  252.6  &  16.62  &  14.62  &  11.82  &  11.48  &  10.98  &  10.73  &   3.14  &   56.4~ &  \\
  SCR 0815-3600  &  08 15 15.98 $-$36 00 58.9  &  0.612  &  0.010  &  350.6  &  16.29  &  13.80  &  11.29  &  10.74  &  10.17  &   9.88  &   3.06  &   34.6~ &  \\
  SCR 0818-3110  &  08 18 40.27 $-$31 10 20.4  &  0.842  &  0.008  &  162.6  &  15.74  &  14.80  &  14.52  &  14.92  &  14.73  &  14.83  &  -0.12  & \nodata &  Probable white dwarf \tablenotemark{b} \\
  SCR 0827-3002  &  08 27 40.82 $-$30 02 60.0  &  0.621  &  0.010  &  330.3  &  16.08  &  13.14  &  11.30  &  10.67  &  10.17  &   9.92  &   2.47  &   41.1~ &  \\
  SCR 0829-2951  &  08 29 09.73 $-$29 51 39.2  &  0.570  &  0.010  &  158.3  &  15.99  &  13.44  &  11.54  &  11.04  &  10.56  &  10.32  &   2.40  &   54.8~ &  \\
  SCR 0843-2937  &  08 43 09.45 $-$29 37 30.9  &  0.514  &  0.007  &  145.1  &  16.13  &  13.94  &  11.68  &  10.53  &  10.01  &   9.72  &   3.41  &   28.3~ &  \\
  SCR 0845-3051  &  08 45 51.93 $-$30 51 31.4  &  0.563  &  0.008  &  257.2  &  16.41  &  14.22  &  12.00  &  10.82  &  10.30  &  10.04  &   3.40  &   32.9~ &  \\
  SCR 0847-3046  &  08 47 09.79 $-$30 46 12.7  &  0.590  &  0.007  &  170.7  &  15.40  &  13.46  &  11.10  &  10.39  &   9.91  &   9.60  &   3.07  &   33.8~ &  \\
  SCR 0904-3804  &  09 04 46.52 $-$38 04 07.5  &  0.643  &  0.007  &  145.0  &  16.88  &  14.93  &  13.16  &  12.03  &  11.57  &  11.36  &   2.90  &   80.2~ &  \\
  SCR 0913-1049  &  09 13 54.20 $-$10 49 33.2  &  0.670  &  0.004  &  219.9  & \nodata &  15.19  &  14.08  &  13.38  &  12.86  &  12.67  &   1.81  & [200.4] &  \tablenotemark{a} \\
  SCR 0914-4134  &  09 14 17.43 $-$41 34 38.9  &  0.749  &  0.008  &  312.5  &  16.33  &  13.69  &  10.98  &   9.98  &   9.42  &   9.12  &   3.71  &   18.2~ &  \\
  SCR 0927-4137  &  09 27 07.25 $-$41 37 12.5  &  0.511  &  0.016  &  120.5  &  11.93  &  10.65  &  10.01  &  10.32  &   9.89  &   9.80  &   0.33  & \nodata &  \tablenotemark{b} \\
  SCR 0956-4234  &  09 56 37.01 $-$42 34 27.5  &  0.620  &  0.005  &  146.8  &  17.00  &  14.67  &  11.82  &  10.99  &  10.47  &  10.21  &   3.68  &   33.3~ &  \\
  SCR 1005-4322  &  10 05 03.16 $-$43 22 28.4  &  0.653  &  0.014  &  292.5  &  14.76  &  12.51  &  10.14  &   9.85  &   9.32  &   9.06  &   2.66  &   29.2~ &  \\
  SCR 1053-3858  &  10 53 49.42 $-$38 58 58.7  &  0.622  &  0.006  &  320.1  &  15.85  &  13.79  &  11.81  &  10.91  &  10.43  &  10.13  &   2.88  &   44.7~ &  \\
  SCR 1058-3854  &  10 58 47.18 $-$38 54 15.2  &  0.565  &  0.006  &  284.1  &  15.65  &  14.60  &  12.19  &  11.01  &  10.52  &  10.21  &   3.59  &   42.9~ &  \\
  SCR 1107-4135  &  11 07 55.93 $-$41 35 52.8  &  1.189  &  0.006  &  282.8  &  16.66  &  14.72  &  13.65  &  12.19  &  11.69  &  11.47  &   2.53  &  [95.4] &  \tablenotemark{a} \\
  SCR 1110-3608  &  11 10 29.03 $-$36 08 24.7  &  0.527  &  0.007  &  268.5  &  17.20  &  15.07  &  12.72  &  10.93  &  10.34  &  10.00  &   4.14  &   22.3~ &  \\
  SCR 1125-3834  &  11 25 37.28 $-$38 34 43.2  &  0.586  &  0.006  &  252.1  &  16.04  &  13.80  &  11.66  &  10.09  &   9.51  &   9.19  &   3.71  &   18.1~ &  \\
  SCR 1132-4039  &  11 32 57.92 $-$40 39 21.4  &  0.725  &  0.008  &  296.8  &  15.26  &  13.38  &  11.22  &  10.38  &   9.89  &   9.65  &   3.00  &   35.9~ &  \\
  SCR 1149-4248  &  11 49 31.61 $-$42 48 10.2  &  0.951  &  0.007  &  259.9  &  15.43  &  13.09  &  12.41  &  11.67  &  11.11  &  10.90  &   1.42  &  [99.4] &  \tablenotemark{a} \\
  SCR 1151-4142  &  11 51 07.83 $-$41 42 17.5  &  0.713  &  0.010  &  247.5  & \nodata &  15.65  &  13.03  &  11.51  &  10.99  &  10.68  &   4.14  &   33.2~ &  \\
  SCR 1159-4256  &  11 59 37.69 $-$42 56 39.3  &  0.610  &  0.007  &  219.0  &  14.20  &  11.96  &  10.35  &   9.54  &   8.98  &   8.72  &   2.42  &   26.7~ &  \\
  SCR 1204-4037  &  12 04 15.54 $-$40 37 52.6  &  0.695  &  0.013  &  150.0  &  14.70  &  12.61  &  10.72  &   9.57  &   9.02  &   8.75  &   3.04  &   21.2~ &  \\
  SCR 1214-4603  &  12 14 40.01 $-$46 03 14.4  &  0.750  &  0.005  &  250.8  &  16.80  &  14.53  &  11.60  &  10.32  &   9.75  &   9.44  &   4.21  &   18.0~ &  \\
  SCR 1220-4546  &  12 20 07.98 $-$45 46 18.2  &  0.758  &  0.005  &  286.3  &  16.53  &  14.59  &  13.35  &  12.70  &  12.16  &  11.95  &   1.89  & [150.8] &  \tablenotemark{a} \\
  SCR 1227-4541  &  12 27 46.82 $-$45 41 16.7  &  1.304  &  0.011  &  282.0  &  16.40  &  14.19  &  13.31  &  12.75  &  12.40  &  12.27  &   1.44  & [188.2] &  \tablenotemark{a} \\
  SCR 1230-3411  &  12 30 01.76 $-$34 11 24.2  &  0.527  &  0.007  &  234.9  &  15.29  &  13.18  &  10.92  &   9.34  &   8.77  &   8.44  &   3.84  &   12.6~ &  \\
  SCR 1247-0525  &  12 47 14.74 $-$05 25 13.5  &  0.722  &  0.007  &  319.8  &  15.90  &  13.38  &  10.92  &  10.13  &   9.62  &   9.29  &   3.25  &   24.2~ &  \\
  SCR 1321-3629  &  13 21 14.84 $-$36 29 18.3  &  0.554  &  0.009  &  247.8  &  18.61  &  16.32  &  13.84  &  12.14  &  11.57  &  11.24  &   4.18  &   38.4~ &  \\
  SCR 1327-3551  &  13 27 39.52 $-$35 51 01.5  &  0.535  &  0.007  &  236.0  &  17.01  &  14.69  &  12.57  &  11.13  &  10.60  &  10.33  &   3.56  &   33.3~ &  \\
  SCR 1400-3935  &  14 00 32.30 $-$39 35 29.4  &  0.507  &  0.006  &  255.7  &  17.45  &  15.44  &  14.20  &  13.47  &  12.90  &  12.66  &   1.97  & [199.7] &  \tablenotemark{a} \\
  SCR 1412-3941  &  14 12 21.14 $-$39 41 33.8  &  0.636  &  0.006  &  240.2  &  16.24  &  14.22  &  12.44  &  10.99  &  10.43  &  10.18  &   3.23  &   37.6~ &  \\
  SCR 1437-4002  &  14 37 21.41 $-$40 02 50.9  &  0.525  &  0.012  &  230.1  &  16.18  &  13.87  &  12.36  &  10.79  &  10.21  &   9.90  &   3.08  &   32.3~ &  \\
  SCR 1455-3914  &  14 55 51.60 $-$39 14 33.2  &  0.798  &  0.012  &  266.4  &  16.61  &  14.52  &  13.57  &  12.50  &  11.98  &  11.79  &   2.02  & [130.8] &  \tablenotemark{a} \\
  SCR 1457-4705  &  14 57 05.34 $-$47 05 26.4  &  0.517  &  0.008  &  226.4  &  16.93  &  15.23  &  14.48  &  13.53  &  12.96  &  12.82  &   1.70  & [215.1] &  \tablenotemark{a} \\
  SCR 1505-4620  &  15 05 27.33 $-$46 20 16.2  &  0.517  &  0.011  &  239.8  &  15.80  &  13.74  &  12.02  &  11.07  &  10.51  &  10.28  &   2.67  &   51.2~ &  \\
  SCR 1511-3403  &  15 11 38.62 $-$34 03 16.6  &  0.561  &  0.006  &  202.9  &  16.04  &  14.05  &  12.09  &  10.05  &   9.42  &   9.13  &   4.00  &   16.1~ &  \\
  SCR 1533-3634  &  15 33 27.70 $-$36 34 02.6  &  0.555  &  0.006  &  237.2  &  16.08  &  14.62  &  13.37  &  11.54  &  10.99  &  10.76  &   3.08  &   58.9~ &  \\
  SCR 1601-3421  &  16 01 55.72 $-$34 21 57.0  &  0.683  &  0.012  &  118.2  &  17.05  &  15.75  &  13.27  &  10.96  &  10.33  &   9.98  &   4.79  &   20.2~ &  \\
  SCR 1608-4442  &  16 08 43.92 $-$44 42 28.8  &  0.628  &  0.012  &  193.1  &  16.59  &  14.95  &  12.94  &  10.88  &  10.35  &  10.10  &   4.07  &   27.5~ &  \\
  SCR 1608-2913AB&  16 08 45.49 $-$29 13 06.6  &  0.540  &  0.016  &  231.0  &  13.61  &  11.65  &   9.91  &   9.68  &   9.15  &   8.51  &   1.97  &   28.9~ &  \tablenotemark{c} \\
  SCR 1613-3040  &  16 13 53.57 $-$30 40 59.0  &  0.522  &  0.009  &  216.7  &  16.68  &  15.41  &  14.32  &  13.15  &  12.58  &  12.38  &   2.26  & [143.0] &  \tablenotemark{a} \\
  SCR 1637-3203  &  16 37 50.55 $-$32 03 11.5  &  0.587  &  0.007  &  221.6  &  18.42  &  15.76  &  13.74  &  11.70  &  11.10  &  10.82  &   4.06  &   30.3~ &  \\
  SCR 1637-4703  &  16 37 56.52 $-$47 03 45.5  &  0.503  &  0.007  &  215.4  &  16.17  &  13.49  &  14.16  &  10.60  &  10.04  &   9.70  &   2.89  &   20.8~ &  \\
  SCR 1648-2049  &  16 48 23.38 $-$20 49 35.4  &  0.679  &  0.008  &  245.8  &  15.71  &  14.00  &  13.25  &  11.56  &  10.99  &  10.77  &   2.44  &   68.6~ &  \\
  SCR 1738-1057  &  17 38 35.48 $-$10 57 25.3  &  0.510  &  0.004  &  178.3  &  17.24  &  15.80  &  14.12  &  11.64  &  11.11  &  10.90  &   4.16  &   41.0~ &  \\
  SCR 1805-4326  &  18 05 12.34 $-$43 26 06.1  &  0.781  &  0.006  &  160.3  &  17.52  &  15.13  &  12.69  &  11.83  &  11.37  &  11.09  &   3.30  &   56.6~ &  \\
  SCR 1811-4239  &  18 11 17.20 $-$42 39 02.5  &  0.732  &  0.010  &  180.9  &  13.68  &  12.13  &  11.21  &  11.38  &  10.82  &  10.65  &   0.75  & \nodata &  \tablenotemark{d} \\
  SCR 1822-0928  &  18 22 44.35 $-$09 28 20.0  &  0.523  &  0.003  &  196.4  &  17.56  &  15.92  &  13.85  &  11.52  &  10.98  &  10.60  &   4.40  &   29.8~ &  \\
  SCR 1841-4347  &  18 41 09.79 $-$43 47 32.6  &  0.790  &  0.007  &  264.2  &  17.65  &  15.19  &  12.32  &  10.48  &   9.94  &   9.60  &   4.71  &   14.6~ &  \\
  SCR 1847-1922  &  18 47 16.69 $-$19 22 20.8  &  0.626  &  0.011  &  230.7  &  15.36  &  13.08  &  10.94  &   9.91  &   9.38  &   9.09  &   3.17  &   23.0~ &  \\
  SCR 1913-1001  &  19 13 24.60 $-$10 01 46.6  &  0.576  &  0.004  &  211.8  &  16.64  &  14.81  &  13.88  &  12.71  &  12.16  &  11.93  &   2.10  & [138.9] &  \tablenotemark{a} \\
  SCR 1916-3638  &  19 16 46.56 $-$36 38 05.9  &  1.303  &  0.007  &  184.1  &  18.20  &  15.88  &  14.78  &  13.66  &  13.12  &  12.95  &   2.22  & [199.2] &  \tablenotemark{a} \\
  SCR 1918-4554  &  19 18 29.45 $-$45 54 31.0  &  0.700  &  0.012  &  220.4  &  17.89  &  15.29  &  12.66  &  11.21  &  10.65  &  10.30  &   4.08  &   25.5~ &  \\
  SCR 1924-3356  &  19 24 48.30 $-$33 56 10.3  &  0.549  &  0.013  &  146.2  &  15.48  &  13.77  &  12.71  &  12.45  &  11.99  &  11.77  &   1.32  & [149.8] &  \tablenotemark{a} \\
  SCR 1931-0306  &  19 31 04.70 $-$03 06 18.6  &  0.578  &  0.004  &   31.0  &  17.87  &  16.06  & \nodata &  11.15  &  10.56  &  10.23  &   4.91  &   18.0~ &  \\
  SCR 1940-3944  &  19 40 21.31 $-$39 44 10.7  &  0.525  &  0.009  &  167.5  &  15.19  &  13.22  &  11.17  &  10.38  &   9.84  &   9.57  &   2.84  &   35.6~ &  \\
  SCR 2001-4239  &  20 01 16.47 $-$42 39 37.1  &  0.594  &  0.008  &  165.5  &  17.08  &  14.90  &  13.25  &  11.84  &  11.34  &  11.09  &   3.06  &   60.7~ &  \\
  SCR 2007-1915  &  20 07 45.91 $-$19 15 53.7  &  0.629  &  0.011  &  186.3  &  16.00  &  13.79  &  11.77  &  10.88  &  10.38  &  10.14  &   2.91  &   43.2~ &  \\
  SCR 2051-1329  &  20 51 13.57 $-$13 29 16.2  &  0.694  &  0.005  &  103.7  &  17.51  &  15.32  &  13.05  &  11.42  &  10.92  &  10.61  &   3.90  &   33.6~ &  \\
  SCR 2132-3922  &  21 32 29.69 $-$39 22 50.3  &  0.531  &  0.007  &  118.3  &  18.36  &  16.14  &  13.54  &  12.21  &  11.70  &  11.35  &   3.93  &   47.7~ &  \\
  SCR 2200-0240  &  22 00 44.45 $-$02 40 18.9  &  0.676  &  0.008  &  174.2  &  17.16  &  15.30  &  13.94  &  12.51  &  11.98  &  11.74  &   2.79  &   97.5~ &  \\
  SCR 2204-3347  &  22 04 02.28 $-$33 47 38.9  &  1.000  &  0.010  &  152.0  &  16.56  &  14.29  &  12.96  &  12.32  &  11.81  &  11.60  &   1.97  & [120.6] &  \tablenotemark{a} \\
  SCR 2247-1528  &  22 47 13.08 $-$15 28 37.8  &  0.512  &  0.008  &  195.4  &  14.02  &  12.08  &  11.31  &  11.10  &  10.50  &  10.34  &   0.98  &   77.3~ &  \\

\tableline %\vspace{-15pt} \\															     					
\multicolumn{14}{c}{SuperCOSMOS$-$RECONS sample with $\mu$ between 0$\farcs$4 yr$^{-1}$ and 0$\farcs$5 yr$^{-1}$} \\					     					
\tableline %\vspace{-15pt} \\															     					

  SCR 0529-3950  &  05 29 40.95 $-$39 50 25.8  &  0.406  &  0.004  &   57.1  &  16.58  &  14.40  &  13.03  &  12.46  &  11.89  &  11.65  &   1.94  & [124.6] &  \tablenotemark{a} \\
  SCR 0533-3908  &  05 33 10.28 $-$39 08 55.5  &  0.454  &  0.005  &   16.5  &  15.88  &  13.73  &  11.58  &  10.71  &  10.18  &   9.90  &   3.02  &   37.0~ &  \\
  SCR 0615-1812  &  06 15 23.95 $-$18 12 04.8  &  0.486  &  0.006  &  150.9  &  18.06  &  15.85  &  13.69  &  12.36  &  11.82  &  11.56  &   3.49  &   62.4~ &  \\
  SCR 0708-4709  &  07 08 32.04 $-$47 09 30.7  &  0.402  &  0.006  &  115.0  &  14.50  &  12.48  &  11.58  &  11.44  &  10.90  &  10.76  &   1.04  &  [93.7] &  \tablenotemark{a} \\
  SCR 0709-3941  &  07 09 37.06 $-$39 41 52.5  &  0.426  &  0.004  &  190.4  &  16.55  &  14.21  &  12.42  &  11.77  &  11.21  &  10.99  &   2.44  &   74.8~ &  \\
  SCR 0709-4648  &  07 09 37.34 $-$46 48 58.6  &  0.413  &  0.006  &   10.4  &  16.03  &  13.49  &  12.57  &  12.20  &  11.70  &  11.49  &   1.29  & [131.3] &  \tablenotemark{a} \\
  SCR 0718-4622  &  07 18 12.12 $-$46 22 37.9  &  0.423  &  0.007  &  343.6  &  17.75  &  15.52  &  13.22  &  11.88  &  11.33  &  11.07  &   3.64  &   46.6~ &  \\
  SCR 0727-1421  &  07 27 16.46 $-$14 21 06.3  &  0.413  &  0.004  &  161.8  &  16.29  &  13.99  &  11.56  &  10.93  &  10.38  &  10.11  &   3.06  &   39.8~ &  \\
  SCR 0727-1404  &  07 27 40.71 $-$14 04 59.0  &  0.484  &  0.004  &  141.7  &  16.78  &  14.55  &  11.95  &  11.35  &  10.79  &  10.52  &   3.20  &   46.5~ &  \\
  SCR 0731-0954  &  07 31 37.56 $-$09 54 50.7  &  0.438  &  0.003  &  177.8  &  18.00  &  15.68  &  12.83  &  11.57  &  11.03  &  10.68  &   4.11  &   32.8~ &  \\
  SCR 0736-3024  &  07 36 56.69 $-$30 24 16.3  &  0.424  &  0.013  &  145.7  &  14.76  &  12.06  &   9.46  &   9.36  &   8.79  &   8.49  &   2.70  &   20.2~ &  \\
  SCR 0740-0540  &  07 40 55.60 $-$05 40 37.9  &  0.467  &  0.003  &  151.1  &  18.20  &  16.04  &  14.62  &  13.51  &  12.96  &  12.77  &   2.53  & [166.4] &  \tablenotemark{a} \\     	
  SCR 0742-3012  &  07 42 41.97 $-$30 12 39.5  &  0.418  &  0.006  &  134.1  &  18.34  &  16.38  &  14.20  &  13.00  &  12.49  &  12.26  &   3.38  &   97.7~ &  \\
  SCR 0745-0725  &  07 45 54.24 $-$07 25 56.1  &  0.437  &  0.003  &  162.8  &  16.71  &  14.55  &  12.92  &  12.18  &  11.64  &  11.42  &   2.37  &   96.9~ &  \\
  SCR 0753-2524  &  07 53 56.58 $-$25 24 01.4  &  0.426  &  0.007  &  300.2  &  16.18  &  15.25  &  15.67  &  14.75  &  14.47  &  14.30  &   0.50  & [365.2] &  Probable white dwarf, common proper motion with LTT2976 \tablenotemark{a} \\
  SCR 0754-2338  &  07 54 29.56 $-$23 38 54.5  &  0.480  &  0.006  &  136.8  &  16.92  &  14.80  &  13.52  &  13.35  &  12.86  &  12.69  &   1.45  & [224.3] &  \tablenotemark{a} \\
  SCR 0754-3809  &  07 54 54.86 $-$38 09 37.4  &  0.401  &  0.011  &  351.4  &  16.90  &  14.68  &  11.75  &  10.01  &   9.42  &   9.08  &   4.67  &   12.0~ &  \\
  SCR 0803-1558  &  08 03 30.08 $-$15 58 30.8  &  0.493  &  0.003  &  153.9  &  17.00  &  14.82  &  12.87  &  12.24  &  11.74  &  11.50  &   2.58  &   94.1~ &  \\
  SCR 0804-1256  &  08 04 48.41 $-$12 56 29.6  &  0.480  &  0.003  &  164.0  &  17.99  &  15.66  &  14.22  &  13.58  &  13.05  &  12.78  &   2.08  & [195.0] &  \tablenotemark{a} \\
  SCR 0816-2247  &  08 16 42.32 $-$22 47 39.8  &  0.418  &  0.006  &  138.4  &  18.51  &  16.26  &  13.40  &  12.75  &  12.21  &  11.90  &   3.51  &   78.5~ &  \\
  SCR 0823-4444  &  08 23 03.57 $-$44 44 50.2  &  0.414  &  0.007  &  308.1  &  16.82  &  14.75  &  12.48  &  11.61  &  11.06  &  10.81  &   3.14  &   54.5~ &  \\
  SCR 0829-3855  &  08 29 23.24 $-$38 55 54.3  &  0.407  &  0.005  &  328.6  &  17.26  &  15.43  &  13.15  &  11.58  &  10.95  &  10.67  &   3.85  &   36.8~ &  \\
  SCR 0835-3400  &  08 35 31.73 $-$34 00 37.4  &  0.448  &  0.012  &  190.1  &  15.31  &  12.76  &  10.25  &   9.90  &   9.37  &   9.08  &   2.86  &   25.9~ &  \\
  SCR 0837-4639  &  08 37 15.70 $-$46 39 50.2  &  0.447  &  0.008  &  303.5  &  16.21  &  13.89  &  12.86  &  12.20  &  11.65  &  11.44  &   1.69  & [119.5] &  \tablenotemark{a} \\
  SCR 0849-3138  &  08 49 38.93 $-$31 38 22.6  &  0.405  &  0.006  &  344.2  &  16.55  &  14.57  &  12.84  &  11.69  &  11.16  &  10.91  &   2.88  &   63.3~ &  \\
  SCR 0917-3849  &  09 17 13.65 $-$38 49 35.8  &  0.484  &  0.007  &  356.7  &  16.62  &  14.37  &  12.28  &  11.56  &  11.12  &  10.80  &   2.81  &   61.3~ &  \\
  SCR 1001-2257  &  10 01 06.62 $-$22 57 04.6  &  0.426  &  0.006  &  142.3  &  17.99  &  15.88  &  14.69  &  14.28  &  13.74  &  13.53  &   1.60  & [328.5] &  \tablenotemark{a} \\
  SCR 1014-4428  &  10 14 40.77 $-$44 28 01.2  &  0.409  &  0.008  &  192.9  &  18.09  &  15.71  &  13.01  &  12.30  &  11.84  &  11.51  &   3.41  &   67.2~ &  \\
  SCR 1109-4631  &  11 09 28.32 $-$46 31 09.9  &  0.467  &  0.005  &  279.2  &  18.17  &  16.18  &  14.00  &  12.35  &  11.88  &  11.55  &   3.83  &   55.5~ &  \\
  SCR 1117-3202  &  11 17 29.31 $-$32 02 09.8  &  0.448  &  0.013  &  204.8  &  15.90  &  13.58  &  11.33  &  10.34  &   9.76  &   9.48  &   3.24  &   26.3~ &  \\
  SCR 1151-4624  &  11 51 01.63 $-$46 24 12.0  &  0.441  &  0.008  &  118.8  &  16.17  &  13.62  &  11.68  &  10.85  &  10.33  &  10.04  &   2.77  &   40.6~ &  \\
  SCR 1157-0149  &  11 57 45.56 $-$01 49 02.4  &  0.451  &  0.008  &  116.4  &  17.29  &  15.13  &  12.62  &  10.90  &  10.35  &  10.02  &   4.23  &   22.2~ &  \\
  SCR 1206-3500  &  12 06 58.52 $-$35 00 52.2  &  0.422  &  0.007  &  229.3  &  15.55  &  13.46  &  11.19  &  10.01  &   9.40  &   9.13  &   3.45  &   21.0~ &  \\
  SCR 1208-3723  &  12 08 51.06 $-$37 23 27.6  &  0.420  &  0.006  &  140.5  &  16.16  &  13.94  &  11.77  &  10.62  &  10.08  &   9.78  &   3.32  &   29.7~ &  \\
  SCR 1223-3654  &  12 23 11.19 $-$36 54 58.5  &  0.461  &  0.006  &  279.4  &  16.60  &  14.53  &  12.45  &  10.99  &  10.42  &  10.15  &   3.54  &   31.9~ &  \\
  SCR 1235-4527  &  12 35 34.99 $-$45 27 03.6  &  0.485  &  0.011  &  317.3  &  14.97  &  12.72  &  11.07  &  10.57  &  10.04  &   9.76  &   2.15  &   48.2~ &  \\
  SCR 1241-4717  &  12 41 33.13 $-$47 17 05.9  &  0.428  &  0.011  &  257.9  &  16.52  &  14.38  &  13.43  &  12.77  &  12.21  &  12.06  &   1.61  & [166.2] &  \tablenotemark{a} \\
  SCR 1246-1236  &  12 46 00.70 $-$12 36 19.4  &  0.406  &  0.007  &  305.4  &  15.84  &  15.80  &  15.86  &  15.74  &  15.73  &  16.13  &   0.06  & \nodata &  White dwarf; K$_s$ unreliable \tablenotemark{b} \\
  SCR 1251-1232  &  12 51 34.75 $-$12 32 59.9  &  0.450  &  0.006  &  264.8  &  16.89  &  14.83  &  13.04  &  12.18  &  11.67  &  11.44  &   2.65  &   89.6~ &  \\
  SCR 1256-1316  &  12 56 31.55 $-$13 16 07.7  &  0.402  &  0.006  &  249.0  &  18.24  &  16.08  &  13.96  &  12.89  &  12.42  &  12.14  &   3.19  &   97.2~ &  \\
  SCR 1340-4427  &  13 40 20.40 $-$44 27 05.8  &  0.403  &  0.005  &  283.5  &  17.22  &  15.18  &  12.72  &  11.69  &  11.16  &  10.88  &   3.49  &   49.0~ &  \\
  SCR 1342-3544  &  13 42 00.21 $-$35 44 51.6  &  0.488  &  0.006  &  283.4  &  18.12  &  16.02  &  14.23  &  13.31  &  12.80  &  12.52  &   2.71  & [141.6] &  \tablenotemark{a} \\
  SCR 1433-3847  &  14 33 03.37 $-$38 47 00.6  &  0.465  &  0.006  &  256.6  &  18.45  &  16.40  &  15.53  &  14.37  &  13.78  &  13.59  &   2.03  & [295.4] &  \tablenotemark{a} \\
  SCR 1444-3426  &  14 44 06.58 $-$34 26 47.3  &  0.451  &  0.014  &  187.7  &  15.01  &  12.49  &  10.47  &   9.74  &   9.18  &   8.88  &   2.75  &   24.0~ &  \\
  SCR 1450-3742  &  14 50 02.86 $-$37 42 10.1  &  0.449  &  0.017  &  212.2  &  15.41  &  13.23  &  11.30  &   9.95  &   9.37  &   9.07  &   3.28  &   21.2~ &  \\
  SCR 1457-3904  &  14 57 49.06 $-$39 04 51.4  &  0.423  &  0.009  &  196.6  &  17.98  &  15.86  &  14.82  &  13.69  &  13.21  &  12.98  &   2.17  & [215.6] &  \tablenotemark{a} \\
  SCR 1507-3611  &  15 07 50.51 $-$36 11 49.7  &  0.407  &  0.005  &  271.6  &  17.67  &  15.51  &  13.67  &  11.55  &  11.05  &  10.78  &   3.96  &   35.2~ &  \\
  SCR 1510-4259  &  15 10 42.34 $-$42 59 25.4  &  0.430  &  0.008  &  229.0  &  17.18  &  15.09  &  12.46  &  11.19  &  10.60  &  10.35  &   3.90  &   31.2~ &  Common proper motion with CD $-$42 10084 \\
  SCR 1512-4354  &  15 12 52.33 $-$43 54 12.3  &  0.419  &  0.011  &  214.2  &  16.00  &  13.69  &  11.47  &  10.57  &   9.96  &   9.75  &   3.12  &   31.6~ &  \\
  SCR 1529-4238  &  15 29 56.31 $-$42 38 38.9  &  0.447  &  0.015  &  243.2  &  16.75  &  14.72  &  13.00  &  11.53  &  10.96  &  10.68  &   3.19  &   34.0~ &  Common proper motion with L 408-087 \\
  SCR 1532-3622  &  15 32 13.90 $-$36 22 31.0  &  0.438  &  0.007  &  235.4  &  15.48  &  13.50  &  11.96  &  10.10  &   9.54  &   9.28  &   3.40  &   23.0~ &  \\
  SCR 1547-2751  &  15 47 36.68 $-$27 51 20.9  &  0.440  &  0.007  &  156.8  &  16.02  &  14.25  &  12.51  &  11.32  &  10.80  &  10.55  &   2.93  &   55.2~ &  \\
  SCR 1550-4718  &  15 50 55.19 $-$47 18 48.4  &  0.413  &  0.013  &  247.6  &  16.12  &  14.24  &  13.23  &  11.79  &  11.19  &  10.98  &   2.45  &   77.2~ &  \\
  SCR 1559-4442  &  15 59 00.74 $-$44 42 12.3  &  0.434  &  0.012  &  220.0  &  15.97  &  14.81  &  14.04  &  12.78  &  12.16  &  11.98  &   2.03  & [125.4] &  \tablenotemark{a} \\
  SCR 1601-4442  &  16 01 37.48 $-$44 42 01.4  &  0.439  &  0.010  &  240.3  &  17.17  &  15.50  &  13.71  &  11.68  &  11.18  &  10.93  &   3.82  &   44.7~ &  \\
  SCR 1608-4229  &  16 08 34.77 $-$42 29 37.8  &  0.408  &  0.008  &  221.3  &  16.97  &  15.81  &  15.23  &  13.57  &  13.01  &  12.87  &   2.24  & [147.6] &  \tablenotemark{a} \\
  SCR 1621-2810  &  16 21 06.94 $-$28 10 24.5  &  0.465  &  0.007  &  163.9  &  17.48  &  16.01  &  15.64  &  13.80  &  13.26  &  13.03  &   2.21  & [197.3] &  \tablenotemark{a} \\
  SCR 1630-3633  &  16 30 27.29 $-$36 33 56.0  &  0.413  &  0.011  &  249.2  &  15.94  &  14.39  &  11.88  &  10.04  &   9.50  &   9.03  &   4.35  &   14.8~ &  \\
  SCR 1631-2805  &  16 31 33.44 $-$28 05 28.0  &  0.468  &  0.008  &  220.7  &  17.46  &  15.67  &  13.82  &  11.90  &  11.36  &  11.06  &   3.77  &   46.3~ &  \\
  SCR 1634-3112  &  16 34 05.78 $-$31 12 02.4  &  0.420  &  0.010  &  248.1  &  16.92  &  14.69  &  12.94  &  11.47  &  10.94  &  10.70  &   3.22  &   46.3~ &  \\
  SCR 1637-4016  &  16 37 03.35 $-$40 16 00.1  &  0.444  &  0.008  &  234.2  &  16.82  &  16.03  &  14.39  &  12.97  &  12.51  &  12.24  &   3.06  &  128.5~ &  \\
  SCR 1637-3014  &  16 37 57.54 $-$30 14 57.3  &  0.462  &  0.014  &  245.1  &  15.39  &  13.36  &  11.65  &  10.89  &  10.34  &  10.11  &   2.47  &   52.4~ &  \\
  SCR 1800-0431A &  18 00 21.33 $-$04 31 47.8  &  0.402  &  0.004  &  227.4  &  17.26  &  16.40  &  15.93  &  13.10  &  12.50  &  12.29  &   3.30  &   96.1~ &  \\
  SCR 1800-0431B &  18 00 20.05 $-$04 32 01.7  & \nodata & \nodata & \nodata & \nodata & \nodata & \nodata &  13.41  &  12.83  &  12.66  & \nodata & \nodata &  \tablenotemark{e} \\
  SCR 1808-0341  &  18 08 48.40 $-$03 41 54.6  &  0.424  &  0.004  &  197.2  &  16.74  &  16.07  &  15.40  &  11.72  &  11.19  &  11.03  &   4.35  &   49.4~ &  \\
  SCR 1822-4542  &  18 22 58.78 $-$45 42 45.9  &  0.436  &  0.006  &  216.2  &  17.86  &  15.13  &  13.76  &  13.65  &  13.10  &  12.88  &   1.48  & [228.7] &  \tablenotemark{a} \\
  SCR 1832-4217  &  18 32 59.19 $-$42 17 20.3  &  0.466  &  0.009  &  200.7  &  15.43  &  13.64  &  12.54  &  12.19  &  11.57  &  11.38  &   1.45  & [121.6] &  \tablenotemark{a} \\
  SCR 1856-1951  &  18 56 15.32 $-$19 51 19.5  &  0.400  &  0.010  &  201.1  &  15.69  &  13.81  &  13.13  &  13.09  &  12.58  &  12.47  &   0.72  &  227.2~ &  \\
  SCR 1857-4309  &  18 57 33.21 $-$43 09 24.9  &  0.403  &  0.020  &  169.5  &  12.32  &  11.83  &  11.33  &  11.09  &  10.75  &  10.69  &   0.74  & \nodata &  \tablenotemark{d} \\
  SCR 1910-4338  &  19 10 23.58 $-$43 38 37.5  &  0.494  &  0.010  &  177.1  &  18.33  &  16.13  &  13.39  &  11.86  &  11.28  &  10.99  &   4.27  &   34.6~ &  \\
  SCR 1913-2312  &  19 13 06.05 $-$23 12 05.4  &  0.416  &  0.006  &  159.2  &  17.89  &  15.81  &  13.40  &  11.43  &  10.87  &  10.52  &   4.38  &   26.0~ &  \\
  SCR 1918-3323  &  19 18 53.29 $-$33 23 56.8  &  0.437  &  0.010  &  221.0  &  17.20  &  15.05  &  12.41  &  11.37  &  10.88  &  10.58  &   3.68  &   39.3~ &  \\     	
  SCR 1928-3634  &  19 28 33.60 $-$36 34 30.1  &  0.470  &  0.012  &  166.4  &  16.36  &  14.12  &  11.92  &  10.61  &  10.06  &   9.81  &   3.51  &   27.4~ &  \\
  SCR 1959-3631  &  19 59 21.03 $-$36 31 03.9  &  0.436  &  0.014  &  158.1  &  11.58  &   9.44  &   8.40  &   8.24  &   7.62  &   7.41  &   1.20  &   19.8~ &  \\
  SCR 2007-3551  &  20 07 41.36 $-$35 51 46.6  &  0.428  &  0.010  &  225.3  &  17.58  &  15.59  &  12.97  &  11.53  &  10.96  &  10.67  &   4.06  &   33.8~ &  \\
  SCR 2044-4123  &  20 44 27.89 $-$41 23 51.6  &  0.429  &  0.012  &  142.5  &  16.12  &  14.03  &  12.52  &  11.75  &  11.16  &  10.99  &   2.28  &   82.4~ &  \\
  SCR 2046-4321  &  20 46 27.46 $-$43 21 06.3  &  0.405  &  0.011  &  173.4  &  14.70  &  13.63  &  13.23  &  12.83  &  12.40  &  12.30  &   0.80  &  169.4~ &  \\
  SCR 2059-4615  &  20 59 10.74 $-$46 15 19.7  &  0.400  &  0.009  &  123.8  &  16.40  &  14.39  &  12.12  &  11.33  &  10.79  &  10.54  &   3.06  &   50.8~ &  \\
  SCR 2059-4302  &  20 59 23.19 $-$43 02 29.7  &  0.448  &  0.009  &  163.7  &  17.38  &  16.03  &  13.49  &  12.75  &  12.21  &  11.96  &   3.28  &  104.4~ &  \\
  SCR 2123-3653  &  21 23 14.36 $-$36 53 27.2  &  0.446  &  0.010  &  133.7  &  17.44  &  15.56  &  13.38  &  12.34  &  11.86  &  11.58  &   3.22  &   78.5~ &  Common proper motion with LTT 8495 \\

\enddata			     

\tablenotetext{a}{distance likely overestimated because object is a subdwarf or white dwarf candidate (see $\S$ 5.3)}
\tablenotetext{b}{all colors too blue for distance relations}
\tablenotetext{c}{separation 2$\farcs$5 at PA 266.2$^\circ$}
\tablenotetext{d}{no distance estimate reported because only one color was available within the bounds of the relations}
\tablenotetext{e}{not detected during automated search due to faint limit but noticed to be a common proper motion companion during the visual inspection, blended on all four plates}

% EXTRA FOOTNOTES --- NOT YET USED

%\tablenotetext{a}{first reported in Hambly et al.~2004; 38.1 $\pm$ 7.8 pc in Henry et al.~2004}
%\tablenotetext{b}{15.4 $\pm$ 2.6 pc in Henry et al.~2004}
%\tablenotetext{c}{10.8 $\pm$ 2.1 pc in Henry et al.~2004}
%\tablenotetext{d}{17.2 $\pm$ 3.1 pc in Henry et al.~2004}
%\tablenotetext{e}{first reported in Hambly et al.~2004; 9.4 $\pm$ 1.7 pc in Henry et al.~2004}
%\tablenotetext{f}{first reported in Hambly et al.~2004; 4.6 $\pm$ 0.8 pc in Henry et al.~2004}
%\tablenotetext{g}{first reported in Hambly et al.~2004; 37.0 $\pm$ 9.4 pc in Henry et al.~2004}
%\tablenotetext{h}{first reported in Hambly et al.~2004; 17.4 $\pm$ 3.5 pc in Henry et al.~2004; K$_s$ suspect}
%\tablenotetext{i}{separation 2$\arcsec$3 at PA 28$^\circ$}
%\tablenotetext{j}{all colors too blue for distance relations}
%\tablenotetext{k}{7.0 $\pm$ 1.2 pc in Henry et al.~2004, binary with separation $\sim$1$\arcsec$0}
%\tablenotetext{l}{separation 11$\arcsec$0 at PA 16$^\circ$}
%\tablenotetext{m}{not detected during automated search due to faint limit but noticed to be a common proper motion companion during the visual inspection; R is $ESO-R$, $R_{59F}$ blended}

\end{deluxetable}

%%%%%%%%%%%%%%%%%%%%%%%%%%%%%% TABLE3: SUBDWARFS %%%%%%%%%%%%%%%%%%%%%%%%%%%%%

%%\voffset100pt{}
\begin{deluxetable}{lcccl}
\tabletypesize{\footnotesize}
\tablecaption{Red Subdwarf Candidates from Paper XII and this Paper.
\label{subdwarfs}}
\tablewidth{0pt}

\tablehead{%%\vspace{-25pt} \\
           \colhead{Name}&
           \colhead{$R_{59F}$}&
           \colhead{$R_{59F}-J$}&
           \colhead{H$_R$}&
           \colhead{TSN Paper}}
\startdata

SCR0242-5935 &  15.02 &  1.46 &  18.36 &  XII \\
SCR0255-7242 &  15.44 &  1.70 &  18.65 &  XII \\
SCR0406-6735 &  14.98 &  1.45 &  18.90 &  XII \\
SCR0433-7740 &  15.86 &  1.81 &  19.41 &  XII \\
SCR0529-3950 &  14.40 &  1.94 &  17.44 &  XV  \\
SCR0629-6938 &  16.23 &  2.56 &  19.60 &  XII \\
SCR0654-7358 &  16.24 &  2.25 &  19.58 &  XII \\
SCR0658-0655 &  14.68 &  2.36 &  18.47 &  XV  \\
SCR0701-0655 &  15.75 &  2.02 &  19.58 &  XV  \\
SCR0708-4709 &  12.48 &  1.04 &  15.50 &  XV  \\
SCR0709-4648 &  13.49 &  1.30 &  16.57 &  XV  \\
SCR0740-0540 &  16.04 &  2.54 &  19.39 &  XV  \\
SCR0754-2338 &  14.80 &  1.45 &  18.21 &  XV  \\
SCR0804-1256 &  15.66 &  2.08 &  19.07 &  XV  \\
SCR0816-7727 &  14.43 &  1.81 &  18.58 &  XII \\
SCR0837-4639 &  13.89 &  1.70 &  17.14 &  XV  \\
SCR0913-1049 &  15.19 &  1.82 &  19.32 &  XV  \\
SCR1001-2257 &  15.88 &  1.60 &  19.02 &  XV  \\
SCR1107-4135 &  14.72 &  2.53 &  20.09 &  XV  \\
SCR1149-4248 &  13.09 &  1.43 &  17.98 &  XV  \\
SCR1220-4546 &  14.59 &  1.89 &  18.98 &  XV  \\
SCR1227-4541 &  14.19 &  1.44 &  19.77 &  XV  \\
SCR1241-4717 &  14.38 &  1.61 &  17.54 &  XV  \\
SCR1320-7542 &  15.82 &  1.89 &  19.01 &  XII \\
SCR1338-5622 &  14.90 &  1.76 &  18.59 &  XII \\
SCR1342-3544 &  16.02 &  2.71 &  19.46 &  XV  \\
SCR1400-3935 &  15.44 &  1.98 &  18.97 &  XV  \\
SCR1433-3847 &  16.40 &  2.03 &  19.74 &  XV  \\
SCR1442-4810 &  14.33 &  1.35 &  17.86 &  XII \\
SCR1455-3914 &  14.52 &  2.02 &  19.03 &  XV  \\
SCR1457-4705 &  15.23 &  1.70 &  18.79 &  XV  \\
SCR1457-3904 &  15.86 &  2.17 &  18.99 &  XV  \\
SCR1559-4442 &  14.81 &  2.03 &  18.00 &  XV  \\
SCR1608-4229 &  15.81 &  2.24 &  18.86 &  XV  \\
SCR1613-3040 &  15.41 &  2.26 &  19.00 &  XV  \\
SCR1621-2810 &  16.01 &  2.21 &  19.34 &  XV  \\
SCR1627-7337 &  13.89 &  1.23 &  17.10 &  XII \\
SCR1735-7020 &  16.14 &  3.32 &  21.06 &  XII \\
SCR1739-8222 &  15.04 &  2.14 &  18.37 &  XII \\
SCR1740-5646 &  15.90 &  2.07 &  19.16 &  XII \\
SCR1756-5927 &  15.73 &  2.29 &  19.38 &  XII \\
SCR1817-5318 &  13.27 &  1.34 &  17.22 &  XII \\
SCR1822-4542 &  15.13 &  1.48 &  18.33 &  XV  \\
SCR1832-4217 &  13.64 &  1.45 &  16.98 &  XV  \\
SCR1835-8754 &  16.02 &  1.92 &  20.05 &  XII \\
SCR1843-7849 &  15.70 &  2.43 &  20.06 &  XII \\
SCR1913-1001 &  14.81 &  2.10 &  18.61 &  XV  \\
SCR1916-3638 &  15.88 &  2.22 &  21.46 &  XV  \\
SCR1924-3356 &  13.77 &  1.33 &  17.47 &  XV  \\
SCR1926-5218 &  15.22 &  1.68 &  18.69 &  XII \\
SCR1946-4945 &  15.39 &  1.88 &  19.22 &  XII \\
SCR1958-5609 &  15.55 &  2.25 &  19.02 &  XII \\
SCR2018-6606 &  15.76 &  2.08 &  19.08 &  XII \\
SCR2101-5437 &  14.59 &  1.80 &  18.71 &  XII \\
SCR2104-5229 &  15.42 &  1.98 &  18.43 &  XII \\
SCR2109-5226 &  15.97 &  2.22 &  20.46 &  XII \\
SCR2151-8604 &  14.48 &  1.74 &  17.77 &  XII \\
SCR2204-3347 &  14.29 &  1.97 &  19.29 &  XV  \\
SCR2235-7722 &  16.36 &  2.19 &  20.29 &  XII \\
SCR2249-6324 &  16.28 &  1.58 &  19.56 &  XII \\
SCR2305-7729 &  15.73 &  1.91 &  18.89 &  XII \\
SCR2317-5140 &  15.02 &  2.19 &  18.26 &  XII \\
SCR2329-8758 &  14.48 &  1.77 &  17.64 &  XII \\
SCR2335-5020 &  15.17 &  2.03 &  19.27 &  XII \\

\enddata

\end{deluxetable}

\clearpage

%%%%%%%%%%%%%%%%%%%%%%%%%%%%%%%%%%%%%%%%%%%%%%%%%%%%%%%%%%%%%%%%%%%%%%%%%%%%%%
%%%%%%%%%%%%%%%%%%%%% TABLE4: DISTANCE ESTIMATE STATS %%%%%%%%%%%%%%%%%%%%%%%%
%%%%%%%%%%%%%%%%%%%%%%%%%%%%%%%%%%%%%%%%%%%%%%%%%%%%%%%%%%%%%%%%%%%%%%%%%%%%%%

%\voffset100pt{}
\begin{deluxetable}{lccc}
\tabletypesize{\footnotesize}
\tablecaption{Distance Estimate Statistics for New SCR Systems.\tablenotemark{a}
\label{diststats}}
\tablewidth{0pt}

\tablehead{%\vspace{-25pt} \\
           \colhead{Proper motion}&
           \colhead{d $\leq$ 10 pc}&
           \colhead{10 pc $<$ d $\leq$ 25 pc}&
           \colhead{d $>$ 25 pc}}

\startdata
$\mu$ $\geq$ 1$\farcs$0 yr$^{-1}$                     &  2 $+$ 0 &  0 $+$ 0  &   2 $+$ 4   \\
1$\farcs$0 yr$^{-1}$ $>$ $\mu$ $\geq$ 0$\farcs$8 yr$^{-1}$ &  0 $+$ 0 &  3 $+$ 0  &   2 $+$ 1   \\
0$\farcs$8 yr$^{-1}$ $>$ $\mu$ $\geq$ 0$\farcs$6 yr$^{-1}$ &  0 $+$ 1 &  4 $+$ 7  &  25 $+$ 23  \\
0$\farcs$6 yr$^{-1}$ $>$ $\mu$ $\geq$ 0$\farcs$4 yr$^{-1}$ &  1 $+$ 1 &  8 $+$ 16 &  95 $+$ 93  \\
\tableline					 
Total                                            &  3 $+$ 2 & 15 $+$ 23 & 124 $+$ 121 \\
\enddata

\tablenotetext{a}{excluding white dwarfs and new wide companions;
first number from Paper XII, second number from this paper}

\end{deluxetable}

\end{document}